\documentclass[a4paper]{jpsj3}

\title{Quantum Spin Liquid in Spin $1/2$ $J_1$-$J_2$ Heisenberg Model on
Square Lattice: Many-Variable Variational Monte Carlo Study Combined
with Quantum-Number Projections}

\author{
Satoshi Morita$^1$\thanks{E-mail: morita@issp.u-tokyo.ac.jp},
Ryui Kaneko$^2$,
and
Masatoshi Imada$^3$
}

\inst{
$^1$Institute for Solid State Physics, University of Tokyo,
Kashiwa, Chiba 277-8581, Japan\\
$^2$Institut f\"{u}r Theoretische Physik, Goethe-Universit\"{a}t
Frankfurt, Max-von-Laue-Stra{\ss}e 1, 60438 Frankfurt am Main, Germany\\
$^3$Department of Applied Physics, University of Tokyo,
Bunkyo, Tokyo 113-8656, Japan
}

\abst{The nature of quantum spin liquids is studied for the spin-$1/2$
antiferromagnetic Heisenberg model on a square lattice containing
exchange interactions between nearest-neighbor sites, $J_1$, and those
between next-nearest-neighbor sites, $J_2$.  We perform variational
Monte Carlo simulations together with the quantum-number-projection
technique and clarify the phase diagram in the ground state together
with its excitation spectra.  We obtain the nonmagnetic phase in the
region $0.4< J_2/J_1\le 0.6$ sandwiched by the staggered and stripe
antiferromagnetic (AF) phases.  Our direct calculations of the spin gap
support the notion that the triplet excitation from the singlet ground
state is gapless in the range of $0.4 < J_2/J_1 \le 0.5$, while the
gapped valence-bond-crystal (VBC) phase is stabilized for $0.5 < J_2/J_1
\le 0.6$. The VBC order is likely to have the columnar symmetry with a
spontaneous symmetry breaking of the $C_{4v}$ symmetry.  The power-law
behaviors of the spin-spin and dimer-dimer correlation functions in the
gapless region are consistent with the emergence of the algebraic
quantum-spin-liquid phase (critical phase). The exponent of the spin
correlation $\langle S(0)S(r)\rangle \propto 1/r^{z+\eta}$ at a long
distance $r$ appears to increase from $z+\eta\sim 1$ at $J_2/J_1\sim0.4$
toward the continuous transition to the VBC phase at
$J_1/J_1\sim0.5$. Our results, however, do not fully exclude the
possibility of a direct quantum transition between the staggered AF and
VBC phases with a wide critical region and deconfined criticality.}

\begin{document}
\maketitle

\section{Introduction}
\begin{figure}[h]
 \centering
 \includegraphics[scale=0.35]{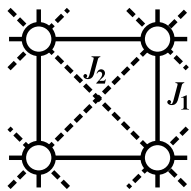}

 \caption{ Lattice structure of the antiferromagnetic $J_1$-$J_2$
 Heisenberg model on a square lattice.  At $J_2=0$, the structure is
 a simple square lattice.  We use the periodic-periodic boundary
 condition.  }

 \label{fig:lattice}
\end{figure}

In the presence of strong geometrical frustration and quantum
fluctuations, insulators without any long range order, {\it i.e.},
quantum spin liquid (SL) states, may appear even at zero temperature.
One of the simplest models proposed for the quantum spin liquid state is
a spin $1/2$ antiferromagnetic $J_1$-$J_2$ Heisenberg model on a square
lattice (Fig.~\ref{fig:lattice}).  The variables $J_1$ and $J_2$ denote
the nearest- and next-nearest-neighbor interactions, respectively.  In
the small-$J_2$ region, just as in the Heisenberg model on a square
lattice, the ground state is widely believed to have the staggered
antiferromagnetic (AF) long-ranged order with a Bragg peak at
$\boldsymbol{q}=(\pi,\pi)$ in the spin structure factor.  On the other
hand, when $J_2$ becomes comparable to $J_1$, the stripe AF long-range
order with Bragg peaks at $\boldsymbol{q}=(0,\pi)$ and $(\pi,0)$ in the
spin structure factor is stabilized. In the intermediate region,
$J_2\sim J_1/2$, geometrical frustration and quantum fluctuations have
been proposed to suppress the long-range magnetic and
valence-bond-crystal (VBC) orders.\cite{prl_dagotto_ed,
prb_poilblanc_ed, prl_capriotti_qmc, prl_capriotti_vmc,
prb_li_bosonic_vmc, prb_hu_vmc, prb_jiang_dmrg, prl_gong_dmrg,
arxiv_wang_tensor_network,arxiv_richter}

There are several high-precision numerical methods of obtaining the
ground states of strongly correlated electron systems.  Among others,
the variational Monte Carlo (VMC) method based on the fermionic
resonating-valence-bond (RVB) state is a powerful tool for examining the
quantum spin liquid states.  More recently, Hu {\it et al.} have
investigated the $J_1$-$J_2$ Heisenberg model by the VMC method together
with the Lanczos technique and reported that the energy gap between the
ground state and the triplet excited state with the total momentum
$\boldsymbol{K}=(\pi, 0)$ closes in the range of $0.48 \leq J_2/J_1 \leq
0.6$.\cite{prb_hu_vmc}

The density matrix renormalization group (DMRG) method is a highly
accurate numerical technique.  It is originally developed for
one-dimensional electron systems and has recently been applied to
two-dimensional ones under the cylindrical boundary condition.  Jiang
{\it et al.} have revisited the ground-state properties of the
$J_1$-$J_2$ Heisenberg model by using the DMRG
method\cite{prb_jiang_dmrg}.  They have reported a spin-gapped quantum
spin liquid phase in the range of $0.41 \leq J_2/J_1 \leq 0.62$.  The
quantum spin liquid state is characterized by the absence of long-range
magnetic and dimer orders.  In contrast to these results, Gong {\it et
al.} showed a gapless region without any magnetic and VBC orders in the
range of $0.44< J_2/J_1 < 0.5$ using DMRG with ${\rm SU}(2)$ spin
rotation symmetry.\cite{prl_gong_dmrg}

In various numerical results, the intermediate region $0.4\lesssim
J_2/J_1 \lesssim 0.6$ has been interpreted as the spin liquid phase with
either gapless~\cite{prl_capriotti_vmc, prb_hu_vmc, prl_gong_dmrg} or
gapful~\cite{prb_li_bosonic_vmc, prb_jiang_dmrg,
arxiv_wang_tensor_network} triplet excitations. However, it has also
been alternatively interpreted by the deconfinement criticality, where a
novel quantum criticality dominated by the deconfinement of magnons
emerges at the critical point between the AF and stripe AF (or VBC)
phases. In this proposal, the spin liquid phase does not exist in the
ground state, but the liquid emerges only at the critical point; in
other words, the parameters away from the critical point always belong
to either of the ordered phases in a strict sense.

It has also been proposed that the intermediate phase contains VBC
phases including the columnar order~\cite{prl_read_vbc, prl_dagotto_ed,
prb_poilblanc_ed, prb_chubukov_dimer, prb_singh_dimer} and plaquette
order.~\cite{prb_zhitomirsky_plaquette, prb_mambrini_vbc,
prl_capriotti_qmc} By using DMRG, Gong {\it et al.}\cite{prl_gong_dmrg}
have reported that a plaquette VBC phase appears for $0.5<J_2/J_1 <
0.61$.

Although both the VMC and DMRG methods can be used to predict the
quantum spin liquid state in the intermediate region of the $J_1$-$J_2$
Heisenberg model, the nature of this state such as the spin gap remains
controversial.  Among all, very recent state-of-the-art studies, one by
VMC~\cite{prb_hu_vmc} and the other two by the DMRG
method~\cite{prb_jiang_dmrg, prl_gong_dmrg} have led to contradictory
conclusions, in terms of the phase diagram and spin liquid properties.
The nature and existence of the quantum spin liquid phase are,
therefore, still under hot debate.

One possible reason for the discrepancy is the inevitable bias existing
in the VMC methods.  As in the case of the calculation by Hu {\it et
al.}, the variational wave functions are often assumed to have a certain
symmetry through the mean-field Hamiltonian~\cite{prb_hu_vmc}.  Another
possible origin of the discrepancy could be the insufficient number of
states kept in the DMRG studies. The limitation of the tractable number
of states also constrains the lattice shape to a cylindrical geometry
and the maximum size of the circumference at most 12 or 14 sites.

To elucidate the origin of the discrepancy, particularly between the VMC
and DMRG results, we perform VMC simulations using improved variational
wave functions that can reproduce both spin-gapped and spin-gapless
states in a unified form.  We employ the many-variable variational Monte
Carlo (mVMC) method~\cite{jpsj_tahara_vmc} for the model of square size
$(L\times L)$ with a periodic boundary condition, which is more
symmetric than the cylindrical boundary condition studied by the DMRG
method and makes the extrapolation to the thermodynamic limit easier.
To reduce biases of the variational wave functions, we introduce a
generalized one-body part of the variational wave functions so that they
can compare both spin-gapped and spin-gapless states on equal footing.
To obtain singlet and triplet excited states, we apply several
quantum-number projections to specify the quantum numbers of the
wave function such as the total spin and momentum, which must be
preserved because they commute with the Hamiltonian.  This procedure not
only enables higher accuracy but also allows us to calculate the energy
gaps and excitation spectra directly.

\begin{figure}[h]
 \centering
 \includegraphics[scale=0.6]{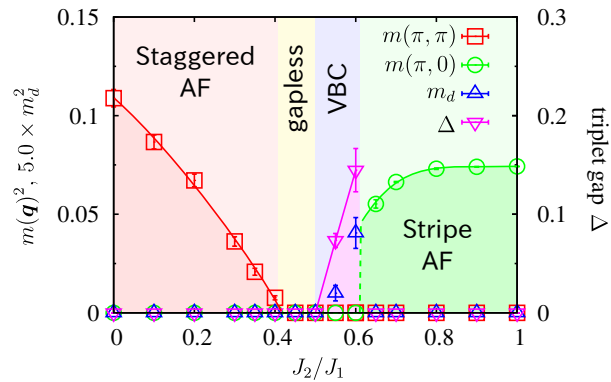}

 \caption{(Color online) Ground-state phase diagram of $J_1$-$J_2$
 Heisenberg model on square lattice obtained in the present study.
 Staggered (stripe) magnetizations are denoted by $m(\boldsymbol{q})$
 with $\boldsymbol{q}=(\pi,\pi)$ ($\boldsymbol{q}=(\pi,0)$).  The dimer
 order parameter $m_d$ is multiplied by $5.0$ and $\Delta$ denotes the
 triplet spin gap. The curves are guides for the eyes. For the
 definitions of $m(\boldsymbol{q})$ and $m_d$, see
 Sect.~\ref{sec:Results}.  }

 \label{fig:phaseDiag}
\end{figure}

Our calculations up to $16\times 16$ sites yield the ground-state phase
diagram after the size extrapolation to the thermodynamic limit, as
shown in Fig.~\ref{fig:phaseDiag}.  The staggered (stripe) AF phase
exists for $J_2/J_1 \leq0.4$ ($J_2>0.6$), and the ground state for $0.4<
J_2/J_1\le 0.6$ has no magnetic order.  In this nonmagnetic region, we
found that the triplet gap closes and becomes gapless in the region
$0.4<J_2/J_1\le 0.5$, while the VBC phase is obtained for $0.5<
J_2/J_1\le 0.6$ with gapful spin-triplet excitations.  We also report
the power-law decay of the spin-spin correlation function in the gapless
region indicating the existence of an algebraic spin-liquid phase in an
extended region.

This paper is organized as follows.  In
Sect.~\ref{sec:Model_and_method}, we first introduce the $J_1$-$J_2$
Heisenberg model and the mVMC method with quantum-number projections.
In Sect.~\ref{sec:Results}, we determine the quantum numbers of the
ground and excited states and report results of the order parameters and
triplet gap. The nature of the nonmagnetic region and the properties of
phase transition points are discussed in Sect.~\ref{sec:Discussion}.
Section \ref{sec:Conclusions} is devoted to the conclusions.

\section{Model and method}
\label{sec:Model_and_method}

We consider the spin $1/2$ antiferromagnetic $J_1$-$J_2$ Heisenberg
model on the square lattice.
The Hamiltonian is given by
\begin{equation}
 H = J_1 \sum_{\left\langle i,j\right\rangle}
     \boldsymbol{S}_i\cdot\boldsymbol{S}_j
   + J_2 \sum_{\langle\!\langle i,j \rangle\!\rangle}
     \boldsymbol{S}_i\cdot\boldsymbol{S}_j,
\end{equation}
where $\left\langle i,j\right\rangle$ and $\langle\!\langle i,j
\rangle\!\rangle$ denote nearest-neighbor and next-nearest-neighbor
sites, respectively; $\boldsymbol{S}_i$ is the spin $1/2$ operator on site $i$.
In the following, we set $J_1=1$ as a unit of
energy.  We calculate the ground state and low-energy
excited states of the model under the periodic boundary conditions.

To obtain the physical properties of the states, we use the mVMC method
with quantum-number projections~\cite{jpsj_tahara_vmc}.  We employ a
fermionic representation of the trial wave functions of the form
\begin{equation}
 \left|\psi\right> =
 \mathcal{P}_{\mathrm{G}}  \mathcal{L}
 \left|\phi_{\mathrm{pair}}\right>,\label{eq:wave_func}
\end{equation}
where $\left|\phi_{\mathrm{pair}}\right>$ and $\mathcal{L}$ denote the
one-body part and quantum number projection, respectively, as we will
detail later.  We introduce the creation (annihilation) operator
$c_{i\sigma}$ ($c_{i\sigma}^\dagger$) of the electron on the site $i$
with spin $\sigma$. The $\alpha$-component of the spin $1/2$ operator
($\alpha=x,y,z$) is represented by
\begin{equation}
 S_i^\alpha = \frac{1}{2} \boldsymbol{c}_i^\dagger
 \sigma_\alpha \boldsymbol{c}_i,
\end{equation}
where $\sigma_\alpha$ denotes the Pauli matrix and
$\boldsymbol{c}_i^\dagger= (c_{i\uparrow}^\dagger,
c_{i\downarrow}^\dagger)$.
The Gutzwiller projection
\begin{equation}
 \mathcal{P}_{\mathrm{G}} = \prod_{i}
 \left(1-n_{i\uparrow}n_{i\downarrow}\right)
\end{equation}
prohibits the double occupation of electrons.

The one-body part is given by a generalized pair wave function defined
as
\begin{equation}
 \left|\phi_{\mathrm{pair}}\right> =
 \left(\sum_{i,j} f_{ij}
 c_{i\uparrow}^{\dagger} c_{j\downarrow}^{\dagger}
 \right)^{N_{\mathrm{s}}/2}\left|0\right>,\label{eq:pair_wave_func}
\end{equation}
where $N_{\mathrm{s}}=L^2$ is the number of sites.  The pairing
amplitudes $f_{ij}$ are taken as a variational parameter depending on $i$
and $j$, and determined by optimization.  This pair wave function can
describe antiferromagnetic, VBC, and spin liquid states on equal footing.
In itinerant systems, in addition to the above variety of Mott
insulators, it can describe metals and superconductors as well.  A
long-range pairing amplitude is necessary to represent a state with spin
correlations decaying with a power law as a function of distance.

In the conventional VMC calculations based on the RVB states, we first
derive the mean-field BCS Hamiltonian and then optimize the order
parameters such as the magnetization and superconducting gap.  The
pairing amplitude of the one-body part is determined by these order
parameters.  Although this approach is simple and intuitive, the wave
functions remain primitive and biased. To reduce such biases, we
directly optimize the pairing amplitude.

In principle, it is better to use a flexible trial wave function without
any constraint on $f_{ij}$.  We however assume a $2\times 2$ sublattice
structure of $f_{ij}$ by assuming the translational invariance in terms
of this sublattice period in order to reduce computational costs.
Namely, we impose the constraint $f_{ij}=f_{kl}$ if
$\boldsymbol{r}_i-\boldsymbol{r}_k
=\boldsymbol{r}_j-\boldsymbol{r}_l=(2n,2m)$ for the arbitrary integers
$n$ and $m$, where $\boldsymbol{r}_i$ is the spatial coordinate of site
$i$.  Therefore, the number of independent variational parameters for
$f_{ij}$ is $4N_{\mathrm{s}}$. This sublattice structure is suitable for
potential candidates of the ground state of this model such as staggered
and stripe AF orders as well as VBC orders.  Preceding works by the
spin-wave analysis\cite{Moreo_incommensurate} and the exact
diagonalization method\cite{prb_poilblanc_ed} did not show any tendency
towards a long-range incommensurate order in the $J_1$-$J_2$ Heisenberg
model.  We confirm that a $2\times 2$ sublattice structure is sufficient
and that larger ones hardly change our results.  In fact, this
constraint itself is justified unless the ground state has a long-range
order with a period longer than the size of the sublattice.  If the
order or fluctuation with the period longer than the $2\times 2$
structure exists in the exact solution, one should expect peaks of the
spin structure factor at the wave vector corresponding to that period
even when symmetry breaking is not allowed by the $2\times 2$ sublattice
structure in our calculation. We will show below that there is no such
tendency, justifying the choice of this sublattice structure.
Meanwhile, we do not impose any constraint between $f_{ij}$ and
$f_{ji}$.

An inner product between the pair wave function and a real-space
electron configuration is represented by a Pfaffian of a skew-symmetric
matrix.  To calculate a Pfaffian, we employ the PFAPACK
library~\cite{acm_wimmer_pfapack} and fast update technique described in
Appendix.

Naive optimizations do not preserve the inherent symmetries of the
finite system.  To restore the symmetries, we take into account
quantum-number projections, which have been successfully used in the
path-integral renormalization group method \cite{prb_mizusaki_pirg} and
Gaussian-basis Monte Carlo method~\cite{jpsj_aimi_gbmc}.  In this study,
we use the spin quantum-number projection, total momentum projection,
and lattice symmetry projection:
\begin{equation}
 \mathcal{L} = \mathcal{L}_S \mathcal{L}_{\boldsymbol{K}}
  \mathcal{L}_{\text{L}}.
\end{equation}
Note that the quantum number projection is commutative with the
Gutzwiller projection.  The spin quantum number projection
$\mathcal{L}_S$ restores the ${\rm SU}(2)$ spin-rotational symmetry by
superposing wave functions rotated in the spin
space\cite{jpsj_tahara_vmc}.

The total momentum projection $\mathcal{L}_{\boldsymbol{K}}$ and the
lattice symmetry projection $\mathcal{L}_{\text{L}}$ restore the
translational symmetry and point group symmetry of the lattice,
respectively.
The former is defined as
\begin{equation}
 \mathcal{L}_{\boldsymbol{K}} \equiv
  \frac{1}{N_{\mathrm{s}}} \sum_{\boldsymbol{R}}
  e^{-i\boldsymbol{K}\cdot\boldsymbol{R}}
  T_{\boldsymbol{R}},
\end{equation}
where $T_{\boldsymbol{R}}$ is a translational operator for shifting all the
spatial coordinates by $\boldsymbol{R}$.
The $2\times 2$ sublattice structure of $f_{ij}$
restricts the total momentum to $\boldsymbol{K}=(0,0)$, $(0,\pi)$,
$(\pi,0)$, or $(\pi,\pi)$.  The point group of the square lattice is
$C_{4v}$ composed of a $\pi/2$ rotation and a reflection along the
vertical axis.  When the total momentum is $\boldsymbol{K}=(0,\pi)$
or $(\pi,0)$, the wave function has the symmetry of $C_{2v}$.
The lattice symmetry projection to an irreducible representation $\beta$
is represented as
\begin{equation}
 \mathcal{L}_{\text{L}}^{(\beta)} = \frac{d_{\beta}}{g}
  \sum_{R}\chi^{(\beta)}_R R,
\end{equation}
where $d_{\beta}$ and $g$ are the dimension of the irreducible
representation and the number of elements $R$ in the point group,
respectively.  The characters $\chi^{(\beta)}_R$ of $C_{4v}$ and
$C_{2v}$ are listed in Table~\ref{table:character}. Here, $C_{4}$ is a
$\pi/2$ rotation and $C_{2}=C_{4}^2$. The reflections along the vertical
axis and the diagonal line are denoted by $\sigma_v$ and $\sigma_d$,
respectively. For $C_{2v}$, $\sigma_x$ is a reflection along the
$x$-axis and $\sigma_y= \sigma_x C_2$.

\begin{table}[h]
 \centering
 \caption{Character tables of $C_{4v}$ and $C_{2v}$}
 \label{table:character}
 \begin{tabular}{c|ccccc} \hline
  $C_{4v}$ & $E$ & $2C_4$ & $C_2$ & $2\sigma_v$ & $2\sigma_d$ \\ \hline
  $A_1$ & 1 & 1 & 1 & 1 & 1 \\
  $A_2$ & 1 & 1 & 1 & -1 & -1 \\
  $B_1$ & 1 & -1 & 1 & 1 & -1 \\
  $B_2$ & 1 & -1 & 1 & -1 & 1 \\
  $E$ & 2 & 0 & -2 & 0 & 0 \\ \hline
 \end{tabular}
 \begin{tabular}{c|cccc} \hline
  $C_{2v}$ & $E$ & $C_2$ & $\sigma_y$ & $\sigma_x$ \\ \hline
  $A_1$ & 1 & 1 & 1 & 1 \\
  $A_2$ & 1 & 1 & -1 & -1 \\
  $B_1$ & 1 & -1 & 1 & -1 \\
  $B_2$ & 1 & -1 & -1 & 1 \\ \hline
 \end{tabular}
\end{table}

A large number of variational parameters are optimized according to the
stochastic reconfiguration method developed by
Sorella~\cite{prb_sorella_sr}.  After confirming the energy convergence,
we calculate the expectation values of the physical quantities and
average them over 10 independent runs to estimate statistical errors.

\section{Results}
\label{sec:Results}

\begin{figure}[h]
 \centering
 \includegraphics[scale=0.38]{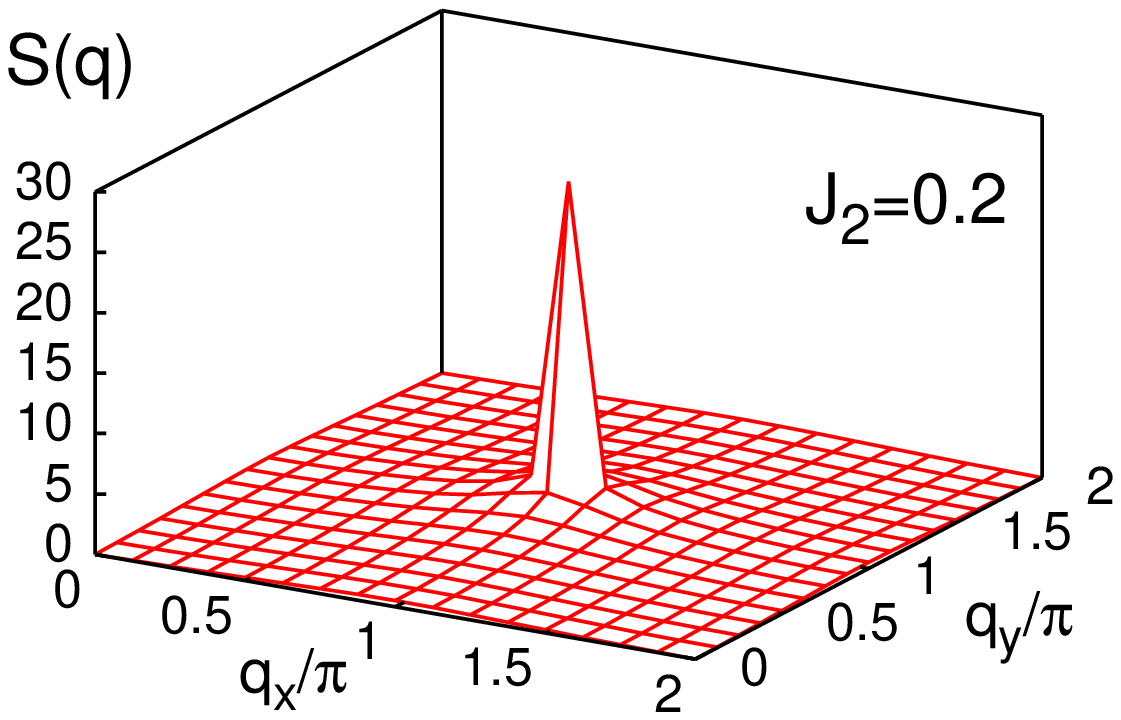}
 \includegraphics[scale=0.38]{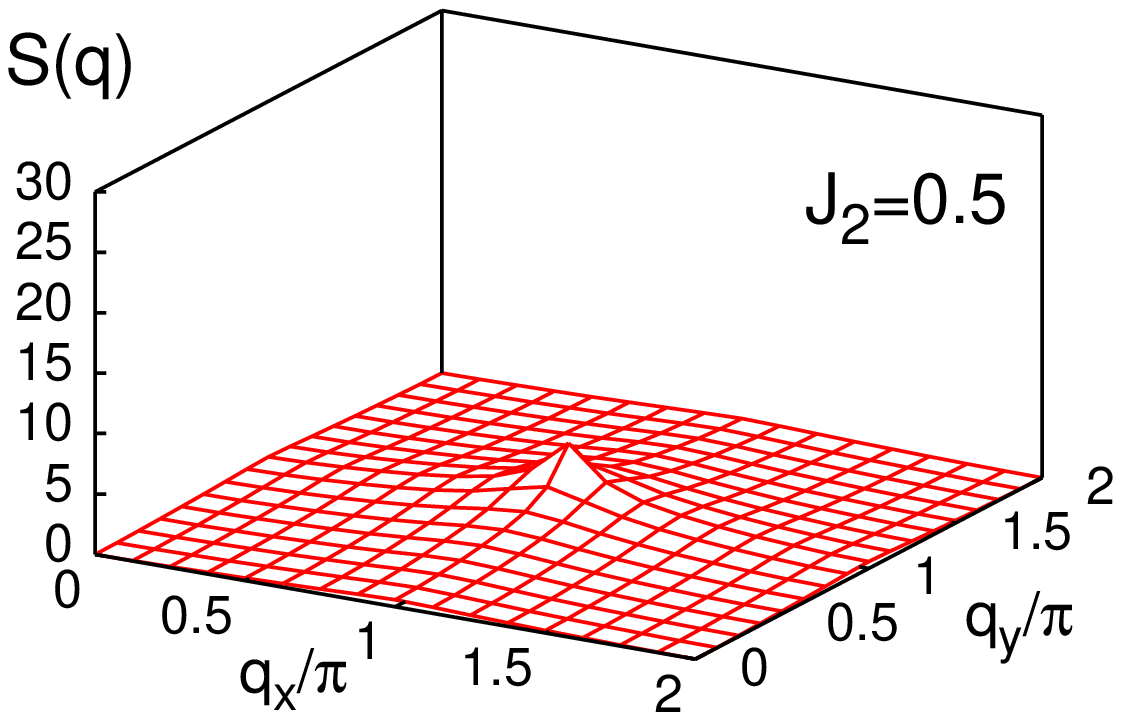}
 \includegraphics[scale=0.38]{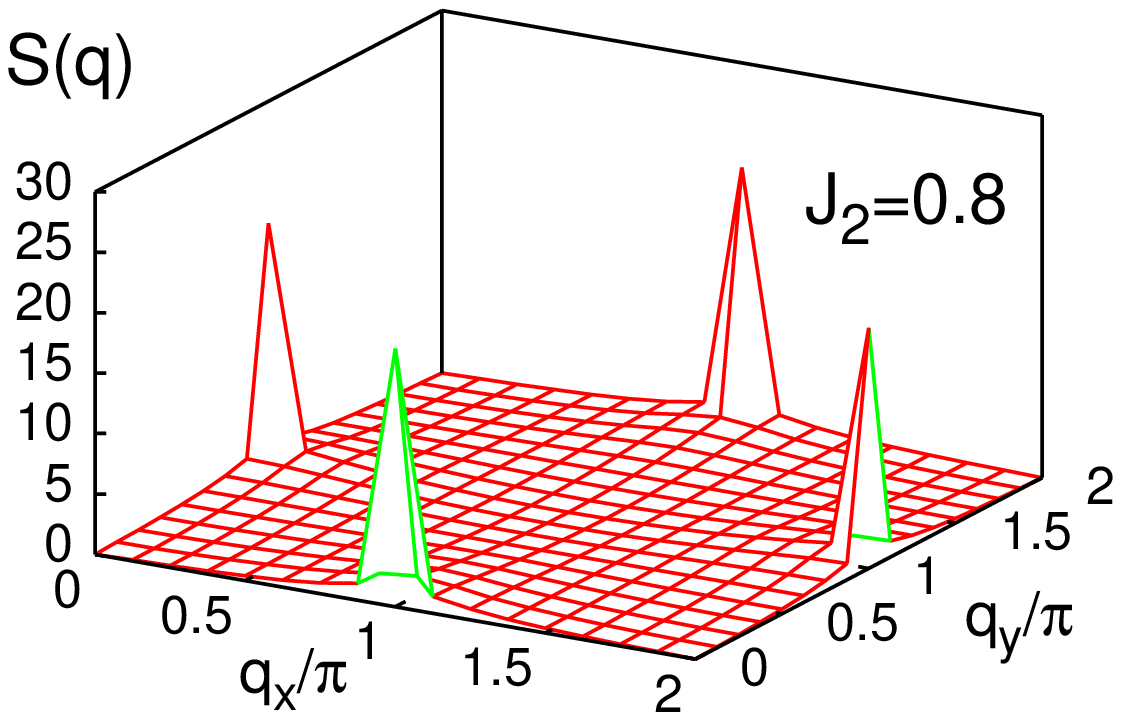}

 \caption{(Color online) Static spin structure factors
 $S(\boldsymbol{q})$ for $16\times 16$ lattice system at $J_2/J_1=0.2$,
 $0.5$, and $0.8$.}

 \label{fig:sq16}
\end{figure}

We first investigate the magnetic property of the ground state to determine
the phase diagram. The spin structure factors, defined as
\begin{equation}
 S(\boldsymbol{q})=\frac{1}{N_{\mathrm{s}}}\sum_{i,j}
  e^{i\boldsymbol{q}\cdot(\boldsymbol{r}_i-\boldsymbol{r}_j)}
  \langle\boldsymbol{S}_i\cdot\boldsymbol{S}_j\rangle,
\end{equation}
are shown in Fig.~\ref{fig:sq16}. For a small $J_2$, the spin structure
factor has a sharp peak at $\boldsymbol{q}=(\pi,\pi)$ corresponding to
the staggered AF state, while it has two sharp peaks at
$\boldsymbol{q}=(\pi,0)$ and $(0,\pi)$ for a large $J_2$ corresponding
to the stripe AF state.  For the intermediate $J_2$, a suppressed peak
appears at $\boldsymbol{q}=(\pi,\pi)$, which signals a nonmagnetic
state.  We emphasize that our trial wave function can represent both
magnetically ordered states because of the direct optimization of the
pairing amplitude.

\begin{figure}[h]
 \centering
 \includegraphics[scale=0.5]{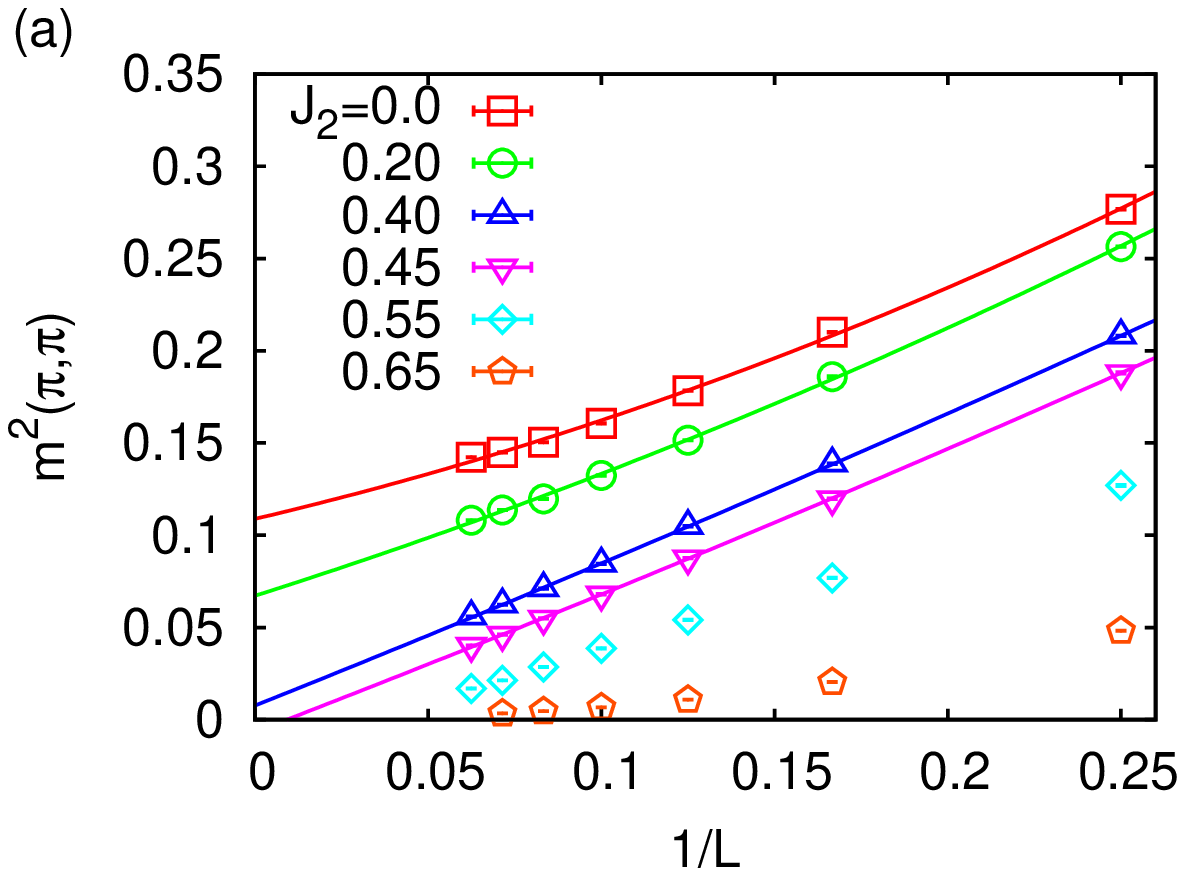}
 \includegraphics[scale=0.5]{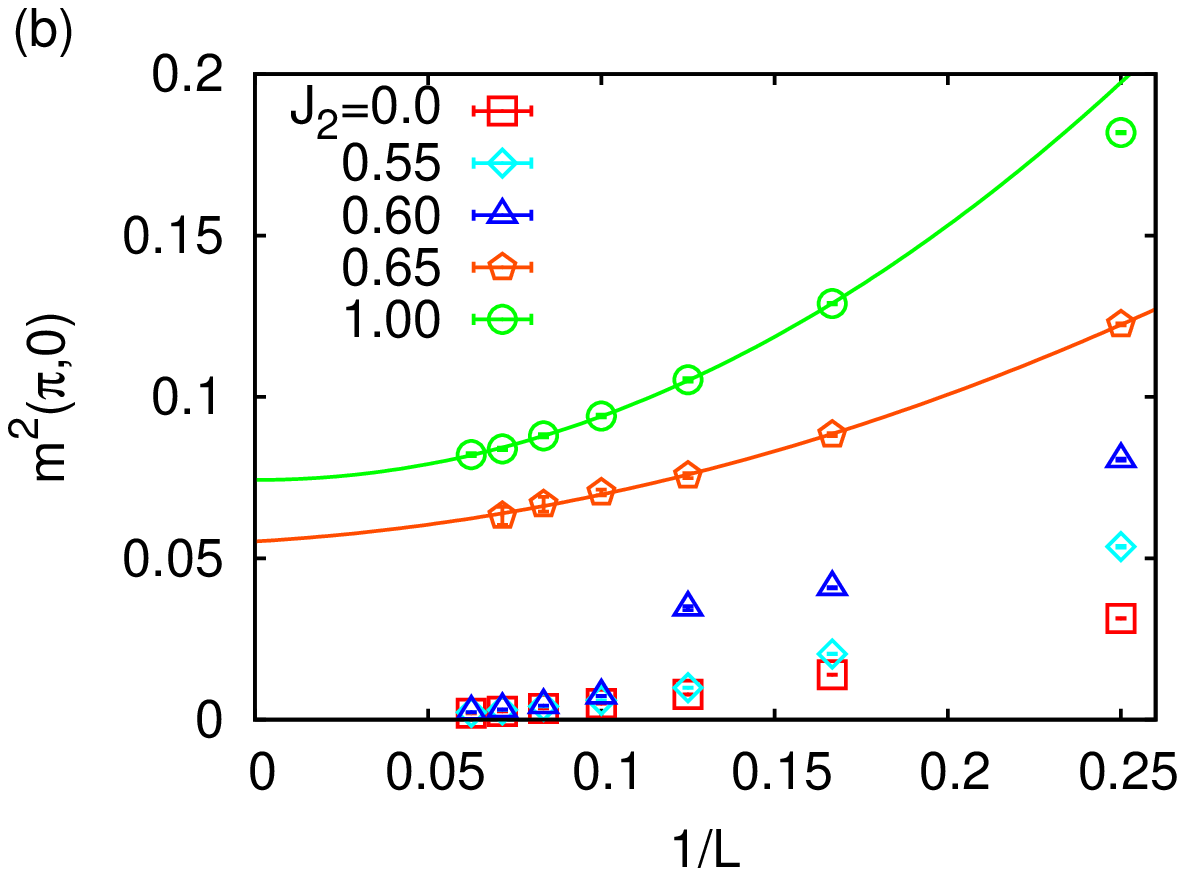}

 \caption{(Color online) Finite-size extrapolation of square
 magnetizations. (a) Staggered AF order parameters with AF wave vector
 $(\pi,\pi)$ and (b) stripe AF order parameters with $(\pi,0)$ are
 fitted by square polynomials of $1/L$. }

 \label{fig:mag_fit}
\end{figure}

\begin{figure}[h]
 \centering
 \includegraphics[scale=0.5]{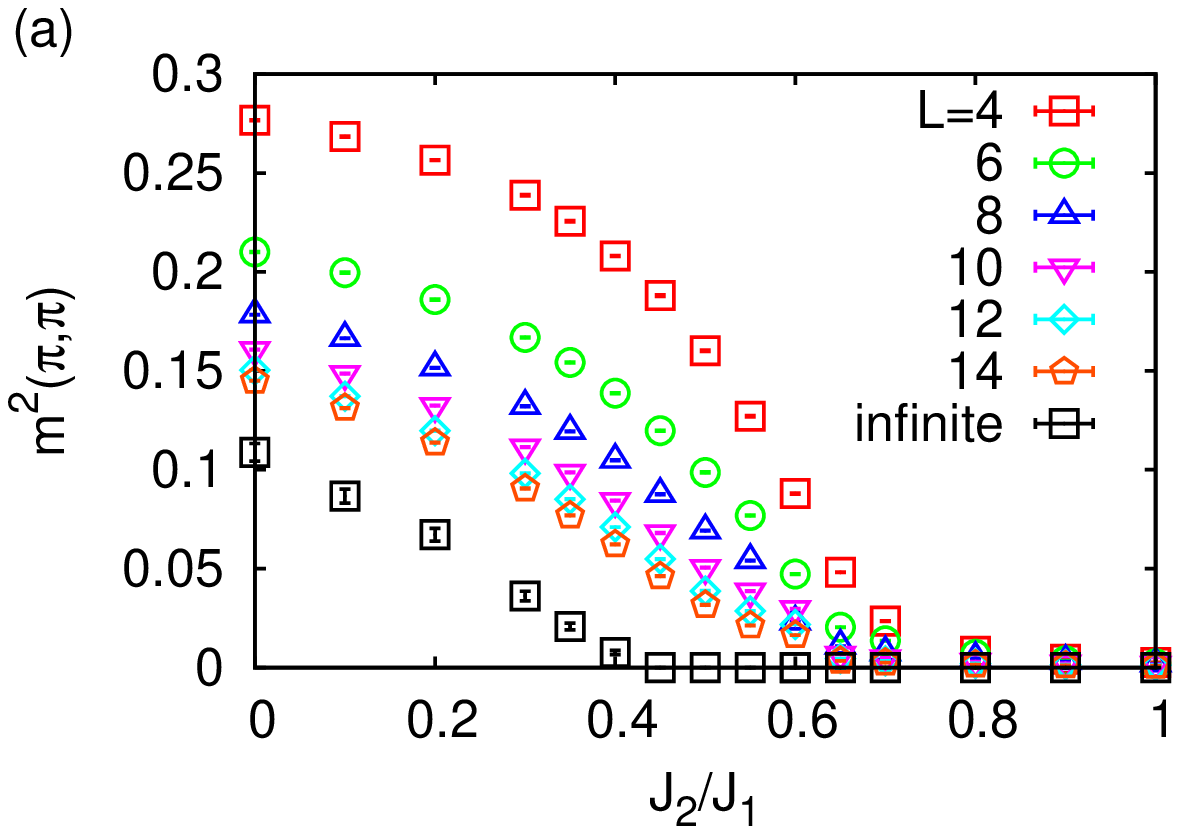}
 \includegraphics[scale=0.5]{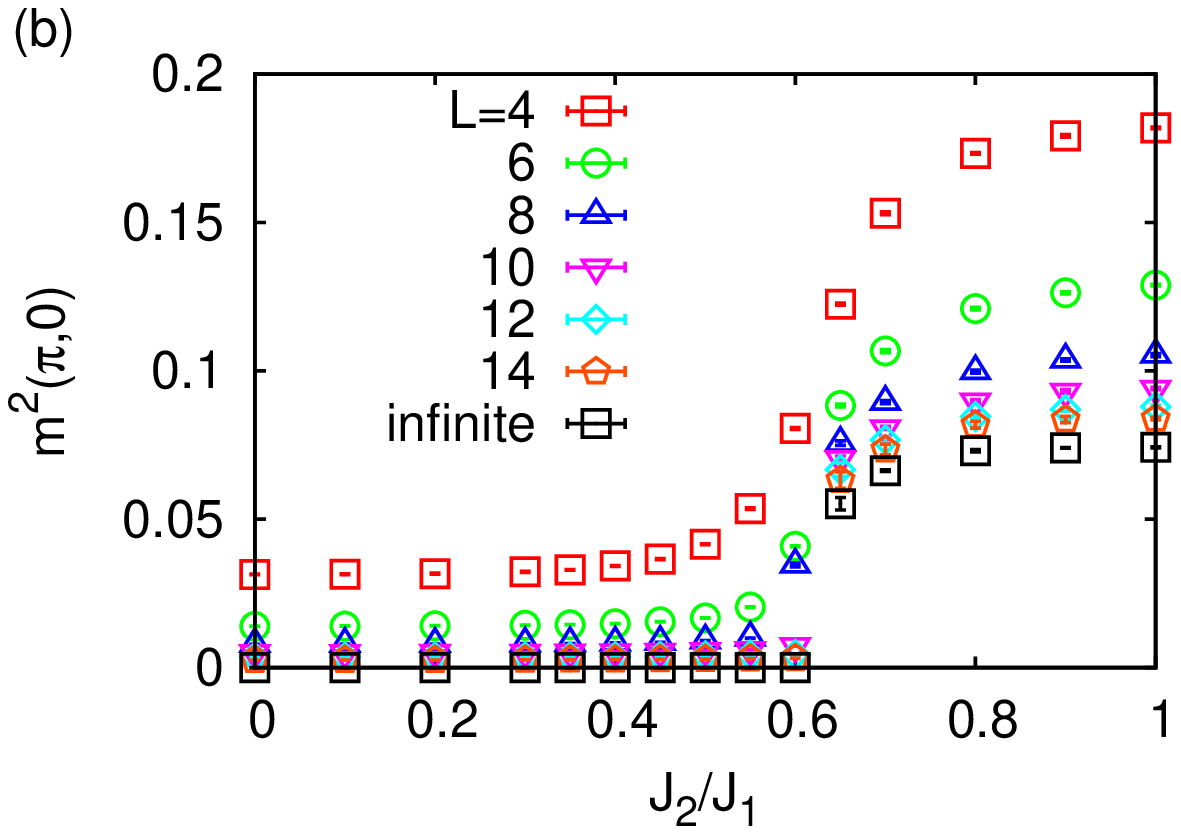}

 \caption{(Color online) (a) Staggered and (b) stripe AF order
 parameters as a function of $J_2/J_1$ plotted for various calculated
 sizes with their extrapolations to the thermodynamic limit.}

 \label{fig:mag_size}
\end{figure}

To analyze the existence of a magnetic long-range order, we extrapolate
the magnetic order parameter
$m(\boldsymbol{q})^2=S(\boldsymbol{q})/N_{\mathrm{s}}$ by fitting the
data with square polynomials of $1/L$ as shown in
Fig.~\ref{fig:mag_fit}.  Note that the spin-wave approximation shows
finite-size correction in power of $1/L$~\cite{PhysRevB.37.2380}.
Finite-size extrapolation shows that the staggered and stripe AF orders
are nonzero for $J_2\leq 0.4$ and $J_2>0.6$, respectively
(Fig.~\ref{fig:mag_size}). Therefore, we conclude that the staggered and
stripe AF phases exist for $J_2\leq 0.4$ and $J_2> 0.6$, respectively
(Fig.~\ref{fig:phaseDiag}). The intermediate region ($0.4 < J_2 \leq
0.6$) has no magnetic order.  The phase boundary and values of
magnetization are in agreement with those in previous studies.

\begin{figure}[h]
 \centering
 \includegraphics[scale=0.5]{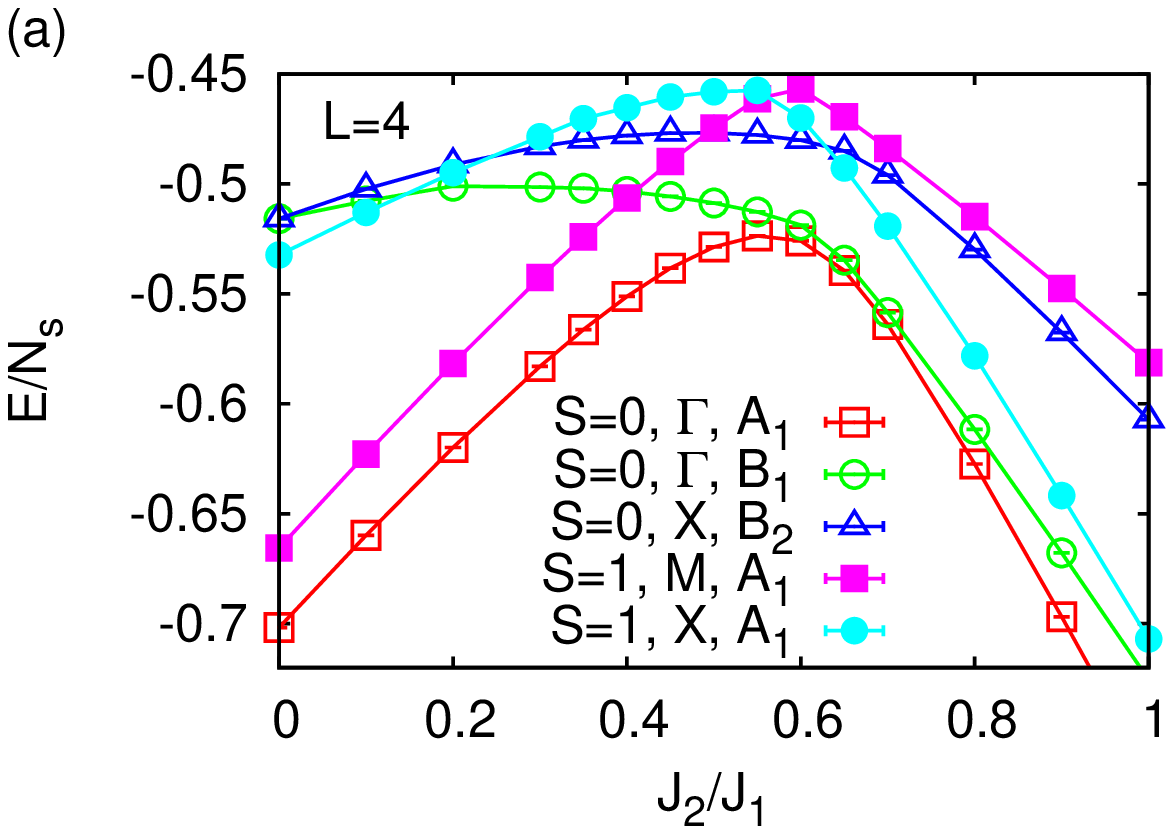}
 \includegraphics[scale=0.5]{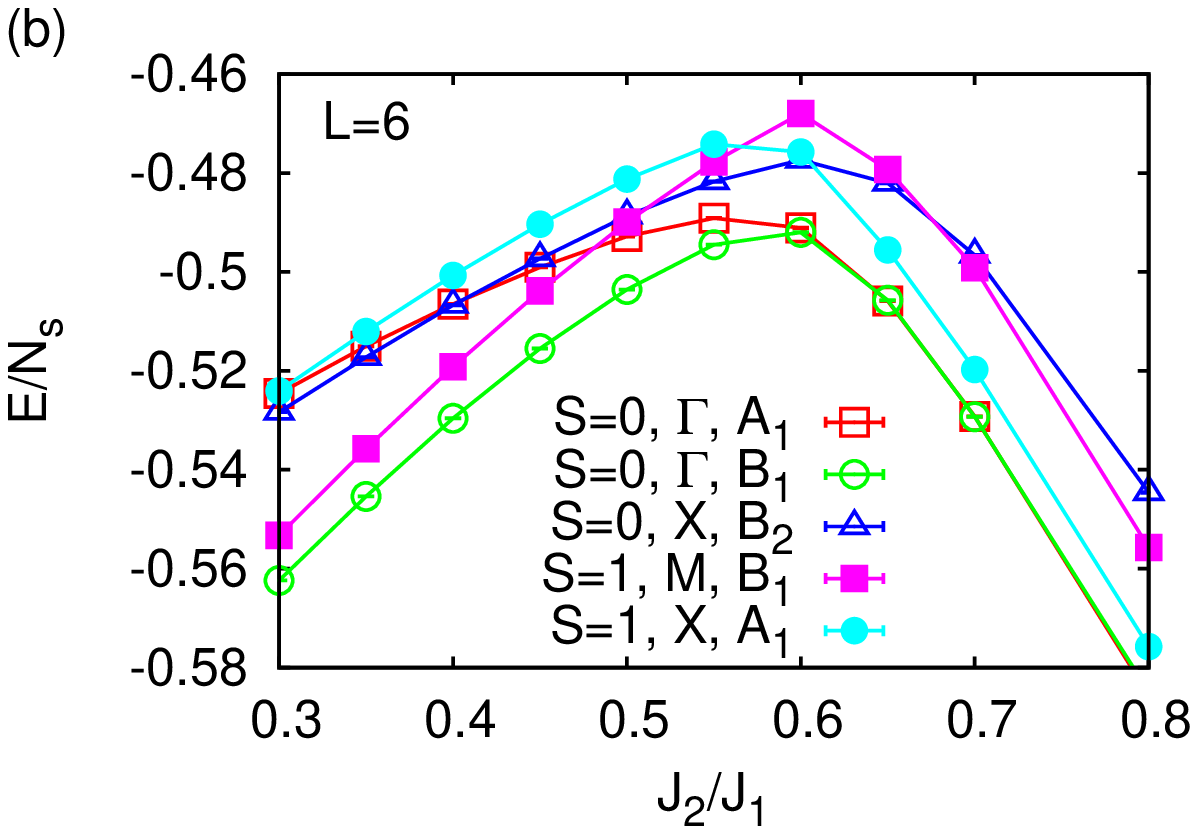}

 \caption{(Color online) Ground-state and excited-state energies for (a)
 $L=4$ and (b) $L=6$ as functions of $J_2$. $A_i$ and $B_i$ ($i=1$ or
 $2$) are irreducible representations of the point group $C_{4v}$ for
 the $\Gamma$ $[\boldsymbol{K}=(0,0)]$ and $M$
 $[\boldsymbol{K}=(\pi,\pi)]$ points and those of $C_{2v}$ for the $X$
 point $[\boldsymbol{K}=(\pi,0)]$.}

 \label{fig:energy}
\end{figure}

\begin{figure}[h]
 \centering
 \includegraphics[scale=0.4]{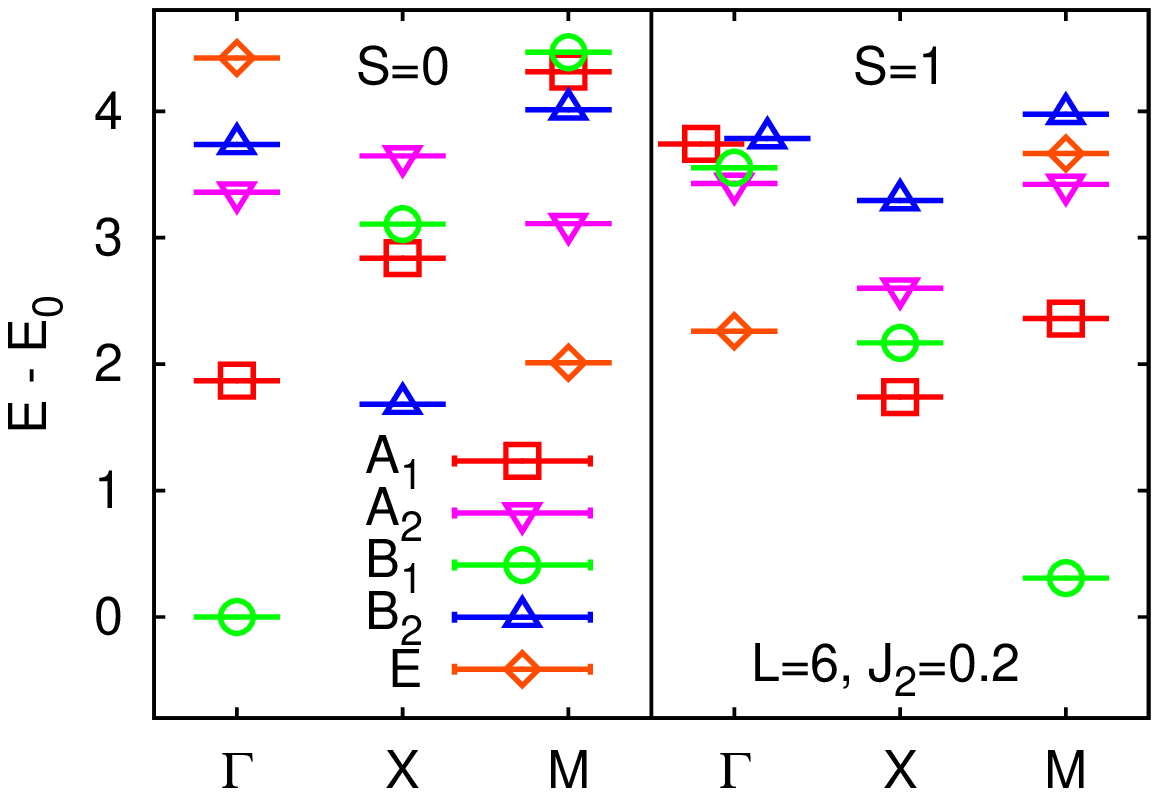}
 \includegraphics[scale=0.4]{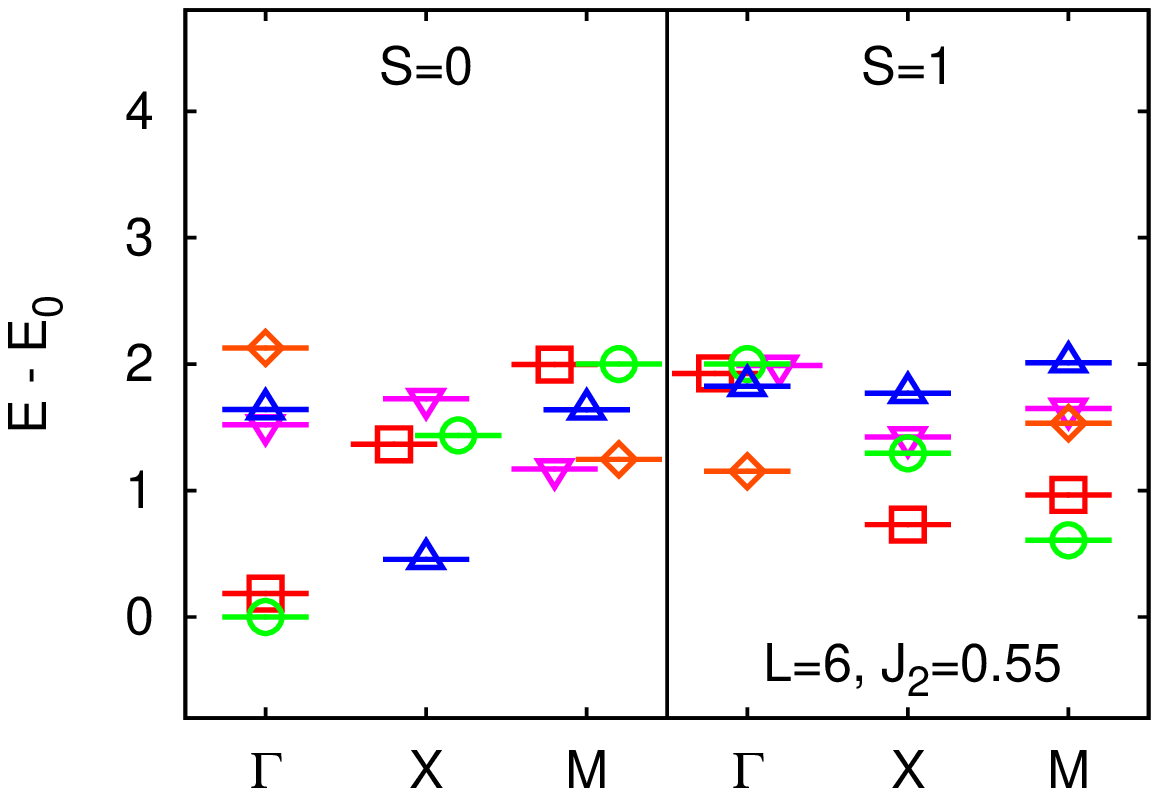}
 \includegraphics[scale=0.4]{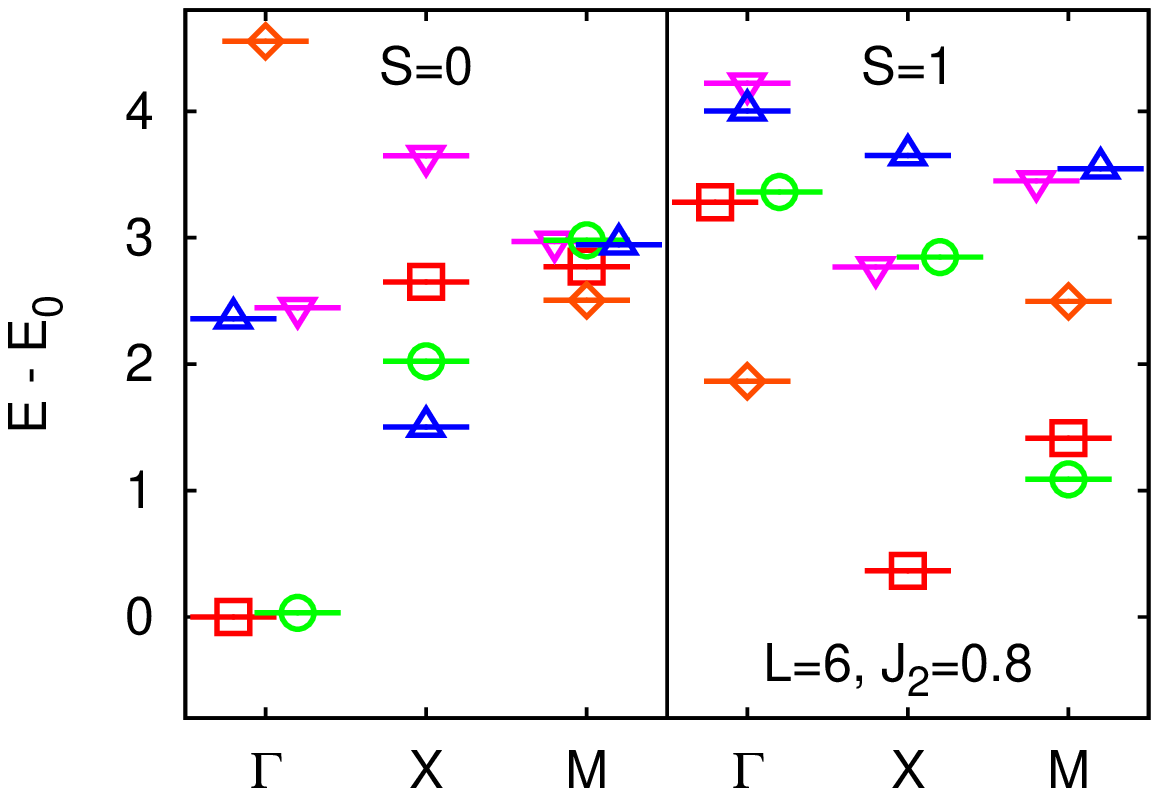}

 \includegraphics[scale=0.4]{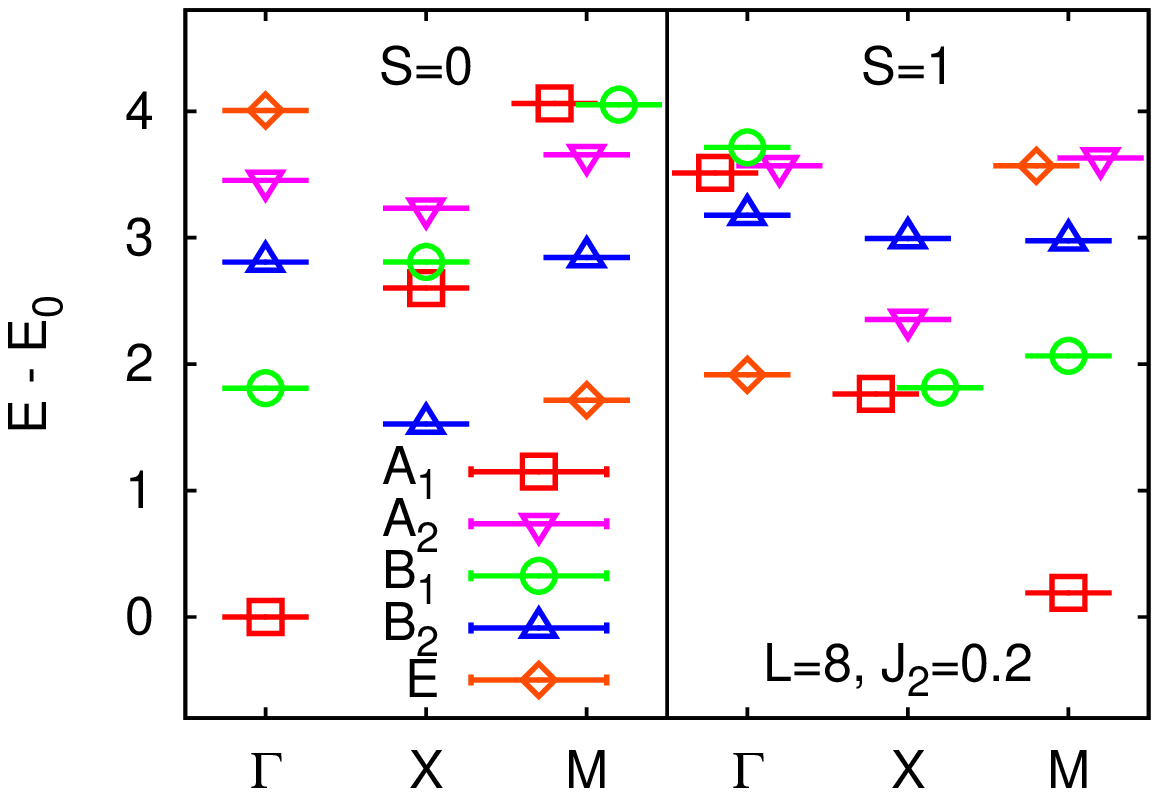}
 \includegraphics[scale=0.4]{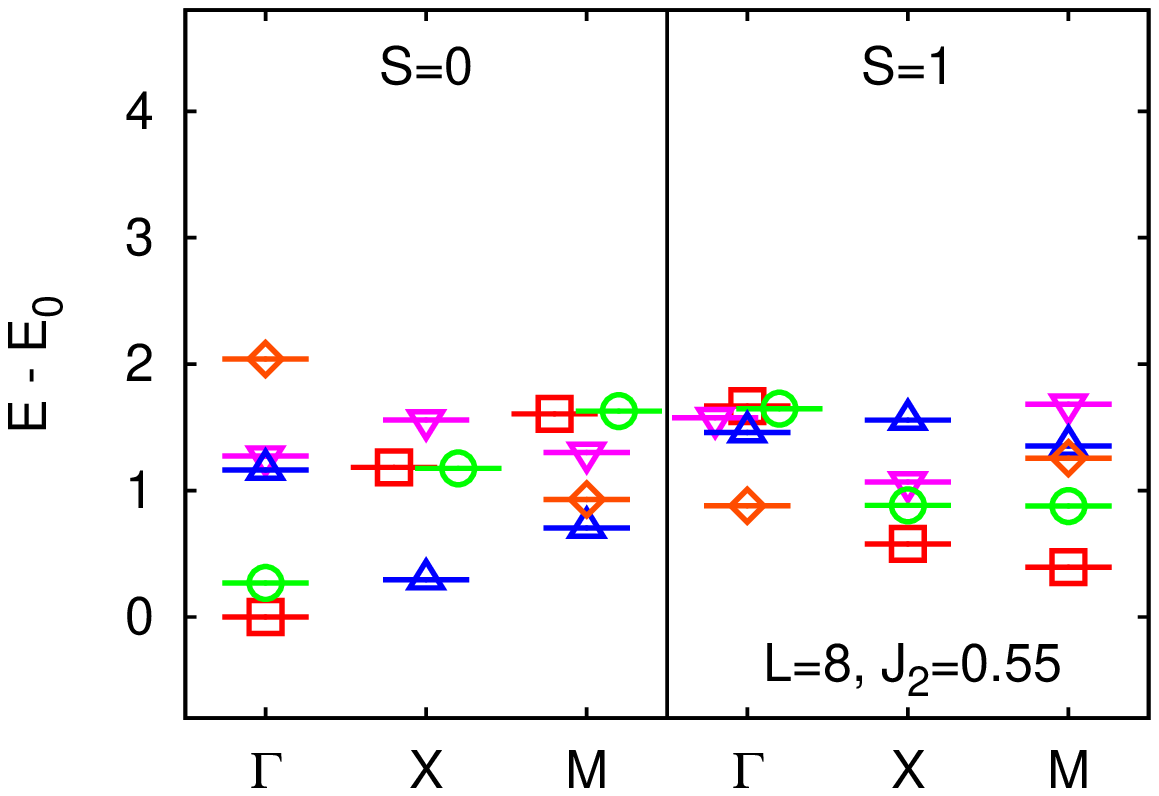}
 \includegraphics[scale=0.4]{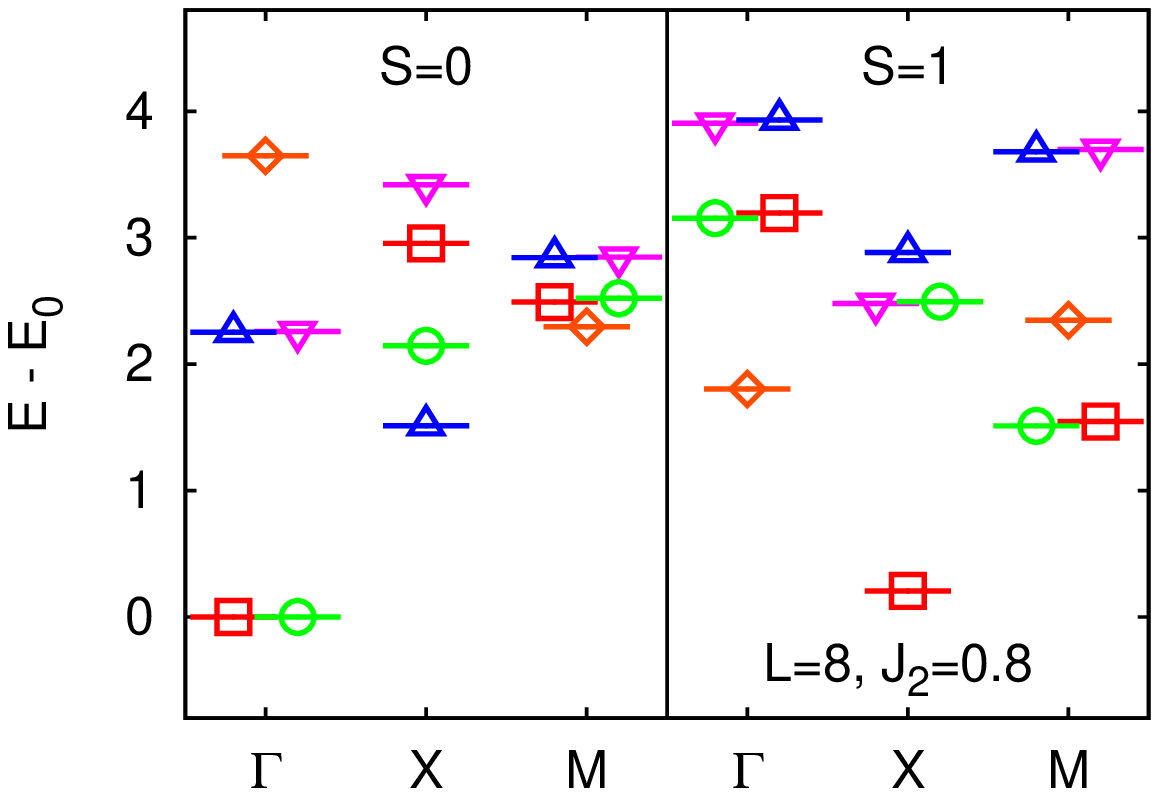}

 \caption{(Color online) Lowest excitation energy with given quantum
 numbers for $L=6$ and $L=8$ with $J=0.2$, $0.55$, and $0.8$. The
 symbols denote irreducible representations of the point group $C_{4v}$
 for $\Gamma$ $[\boldsymbol{K}=(0,0)]$ and $M$
 $[\boldsymbol{K}=(\pi,\pi)]$ and of $C_{2v}$ for $X$ point
 $[\boldsymbol{K}=(\pi,0)]$.  }

 \label{fig:energy_level}
\end{figure}

The quantum numbers of the ground state are determined by calculating
the lowest energy with every combination of total spin ($S=0$ or $1$),
total momentum, and lattice symmetry.  In Figs.~\ref{fig:energy} and
\ref{fig:energy_level}, we show the lowest energy with given quantum
numbers.

The ground state always has a total momentum $\boldsymbol{K}=(0,0)$ and
a total spin $S=0$, and an even parity under the reflection along the
vertical axis.  The parity of $\pi/2$ lattice rotation, however, depends
on the system size.  The ground state of $L=4n$ has an even parity for
all $J_2$ values.  On the other hand, for $L=4n+2$, the energy level
crossing occurs at the transition point $J_2\sim 0.6$ between the
nonmagnetic region and the stripe AF phase. The ground state for a
small (large) $J_2$ is odd (even) under $\pi/2$ rotation.  For
$J_2>0.6$, the ground state with an even parity for $\pi/2$ rotation and
the first singlet excited state with an odd parity eventually degenerate
in the thermodynamic limit.

As shown in Fig.~\ref{fig:energy}, the lowest triplet excited states
cross at the phase boundary between the staggered AF phase and the
nonmagnetic region.  The first triplet excited state for $J_2< 0.6$ has
a total momentum $\boldsymbol{K}=(\pi,\pi)$, while that for $J_2>0.6$
has $\boldsymbol{K}=(\pi,0)$ or $(0,\pi)$, which corresponds to the
staggered and stripe AF wave vectors, respectively.  Both triplet excited
states are even under the reflection along the vertical axis.  For $J_2<
0.6$, the lowest triplet excited state has the same parity of $\pi/2$
rotation as the ground state. Thus, in this region, the triplet gap is
defined as the energy difference between the singlet state with
$\boldsymbol{K}=(0,0)$ and the triplet state with
$\boldsymbol{K}=(\pi,\pi)$.  For $L=4n$ ($4n+2$), both states belong
to the same irreducible representation $A_1$ ($B_1$) of $C_{4v}$.  We
will discuss in the next section how we should understand the symmetry
of the ground state and the excitation spectra.

Next, we investigate the triplet gap to characterize the nonmagnetic
region for $0.4< J_2 \leq 0.6$. As we previously showed, the lowest
triplet state in this region has a total momentum
$\boldsymbol{K}=(\pi,\pi)$ and belongs to the irreducible representation
$A_1$ ($B_1$) for $L=4n$ ($4n+2$).

We consider two types of the system size dependence of the triplet gap:
\begin{equation}
 \Delta(N_{\mathrm{s}}) = \Delta_\infty
  + \frac{a}{N_{\mathrm{s}}}+\frac{b}{N_{\mathrm{s}}^{3/2}},
  \label{eq:gap_fit}
\end{equation}
and
\begin{equation}
 \Delta(N_{\mathrm{s}}) = \Delta_\infty
  + \frac{a'}{N_{\mathrm{s}}^{1/2}}+\frac{b'}{N_{\mathrm{s}}},
  \label{eq:gap_fit2}
\end{equation}
where $\Delta_\infty$ denotes the triplet gap in the thermodynamic limit.
The former scaling form consists with a dispersion relation of the triplet
excitation\cite{prl_white_spingap,prl_santoro_gfmc},
\begin{equation}
 \Delta(\boldsymbol{k})
  = \sqrt{\Delta_\infty^2
  + v^2 (\boldsymbol{k}-\boldsymbol{k}_0)^2},
 \label{eq:dispersion}
\end{equation}
because the nonzero wave vector scales as
$|\boldsymbol{k}-\boldsymbol{k}_0| \sim \pi/L$ in the finite-size
system.  The spin-wave analysis also shows that the leading order of the
triplet gap is scaled by $1/N_{\mathrm{s}}$ for an antiferromagnetically ordered
state in two-dimensional
systems.\cite{prb_neuberger_sw,zpb_hasenfratz_sw} Hence, if the gap is
nonzero or if the AF order exists, this scaling form is more
appropriate. The latter form Eq.~(\ref{eq:gap_fit2}) with
$\Delta_\infty=0$ is justified at the quantum critical point and in the
quantum critical phase with the dynamical critical exponent $z=1$. If we
obtain a negative gap $\Delta_\infty<0$ by size extrapolation, we fit
the parameters again with fixed $\Delta_\infty=0$.

\begin{figure}[h]
 \centering
 \includegraphics[scale=0.5]{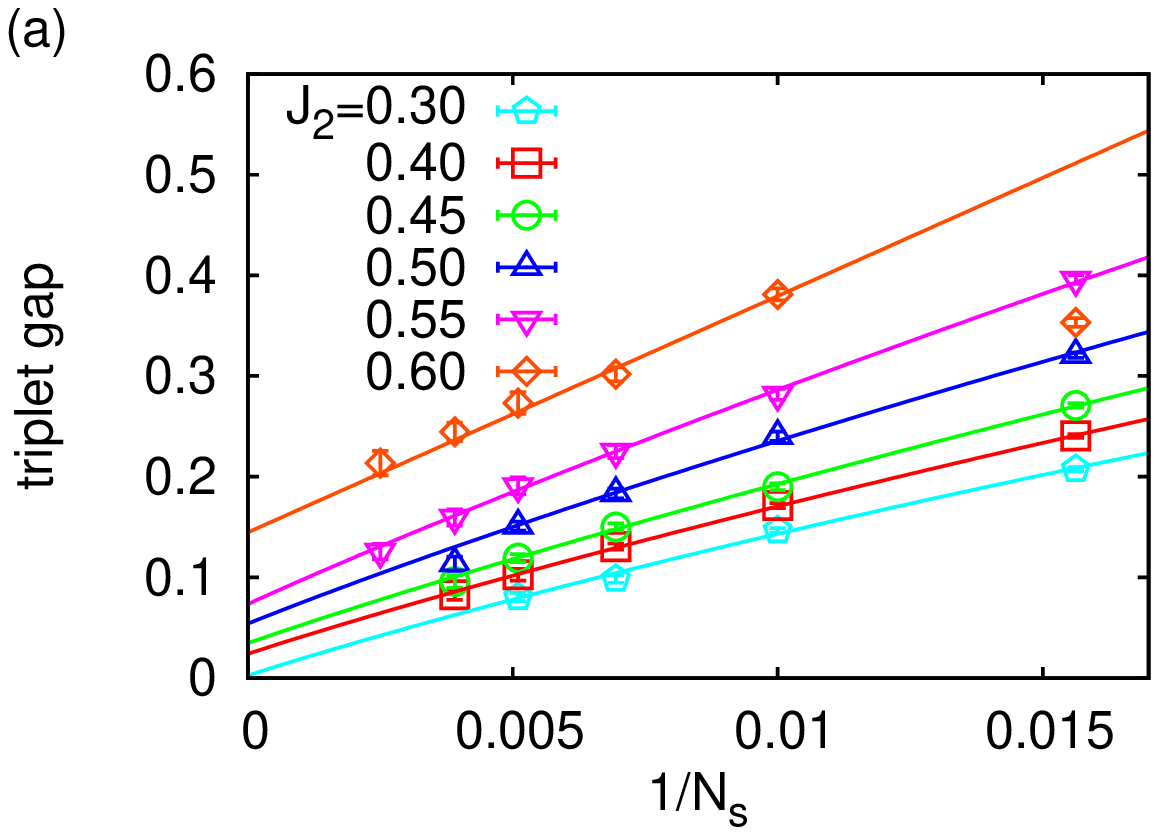}
 \includegraphics[scale=0.5]{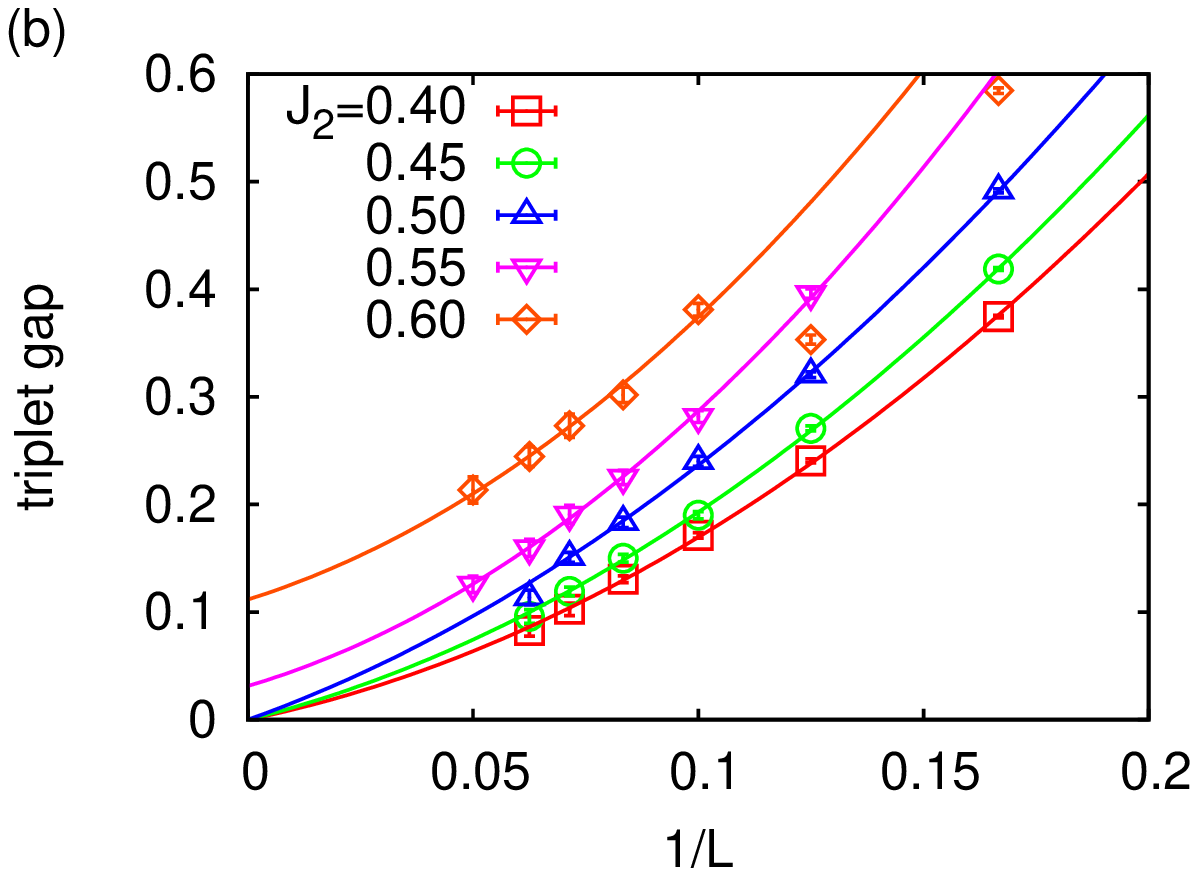}

 \caption{(Color online) Finite-size extrapolation of triplet gap
 $\Delta(L)$.  (a) Triplet gaps are plotted as a function of
 $1/N_{\mathrm{s}}$, and fits of the data to the form
 Eq.~(\ref{eq:gap_fit}) are shown as solid curves. (b) Triplet gaps as a
 function of $1/L$ are fitted by Eq.~(\ref{eq:gap_fit2})}

 \label{fig:gap}
\end{figure}

In Fig.~\ref{fig:gap}, we show the spin gap for various system sizes.
The data fit well the scaling form Eq.~(\ref{eq:gap_fit}) in the whole
range of $J_2$ and we obtain a nonzero spin gap in the thermodynamic
limit for $0.4\leq J_2 \leq 0.6$.  As shown in Fig.~\ref{fig:gap}(b),
however, the spin gap value in the range of $0.4\leq J_2 \leq 0.5$ can
also be fit by the scaling form Eq.~(\ref{eq:gap_fit2}) indicating
vanishing gap.  The triplet gap is concave-downward as a function of
$1/N_{\mathrm{s}}$, which implies that the scaling form
Eq.~(\ref{eq:gap_fit}) may overestimate the triplet gap in the
thermodynamic limit. On the other hand, the convex fitting of the
triplet gap as a function of $1/L$ in Fig.~\ref{fig:gap}(a) may lead to
underestimation. We show below that the triplet excitation is likely to
be gapless for $0.4\leq J_2 \leq 0.5$ making the scaling form in
Fig.~\ref{fig:gap}(b) more appropriate. However, for $J_2>0.5$, even the
scaling form Eq.~(\ref{eq:gap_fit2}) produces a nonzero spin gap. The
gapful triplet excitation for $J_2>0.5$ is consistent with the VBC order
identified below.

\begin{figure}[h]
 \centering
 \includegraphics[scale=0.5]{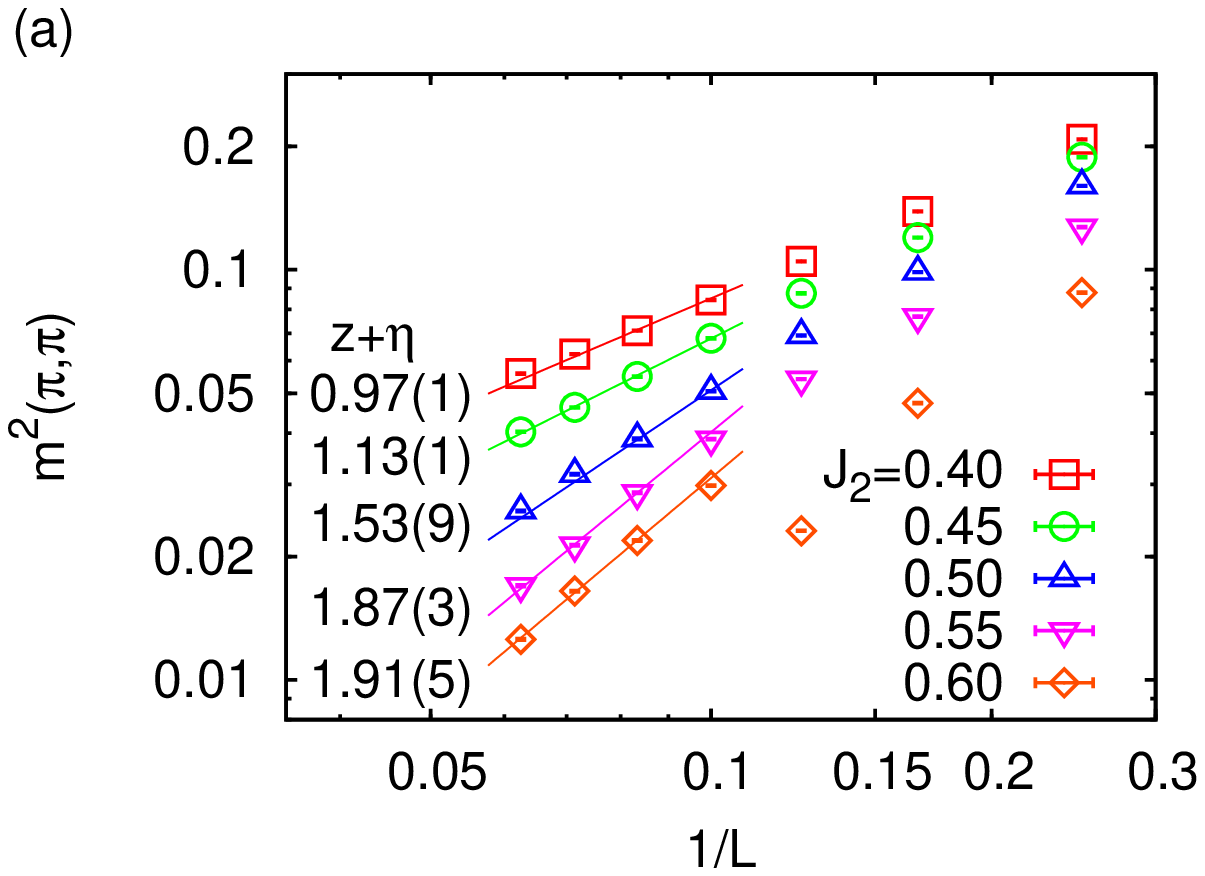}
 \includegraphics[scale=0.5]{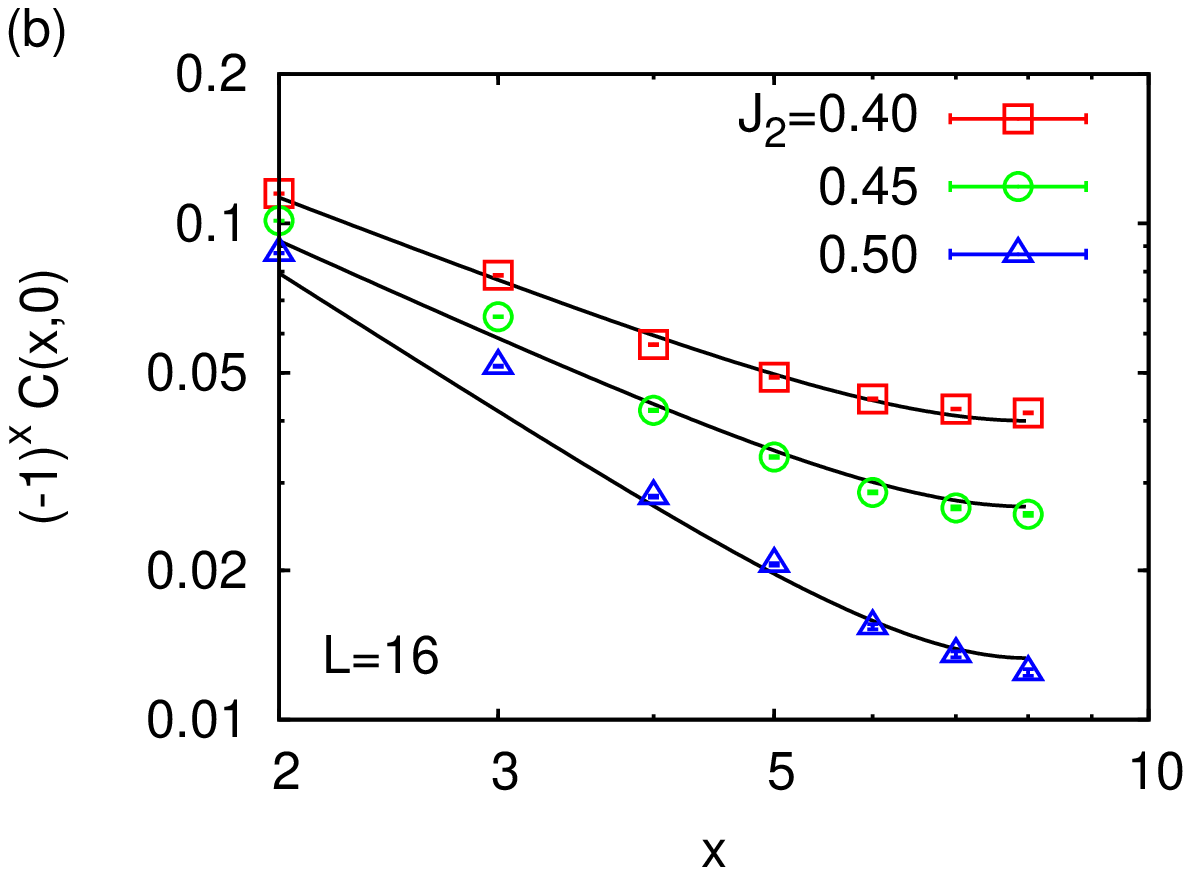}

 \caption{(Color online) (a) Log-log plot of magnetic order
 parameter. The solid lines are obtained by fitting the data with
 $L^{-(z+\eta)}$. The uncertainties in the last digits of the numerical
 data are determined by the fitting and do not take into account
 possible systematic finite-size effects. (b) Spin-spin correlation
 functions along the $x$-axis for $L=16$. The solid lines are
 Eq.~(\ref{eq:spin_fit}) with $z+\eta$ obtained from the left panel.}

 \label{fig:mag_log}
\end{figure}

To further clarify the criticality of the nonmagnetic region
$0.4<J_2\leq 0.5$, we re-examine the size dependence of the magnetic
order parameter $m(L)$. If we assume that the correlation function
decays as $C(r)\propto r^{-(d+z-2+\eta)}$, the peak value of the
structure factor is expected to follow the system size scaling
\begin{equation}
 S(\boldsymbol{q}_{\rm peak}, L)\sim
  \int_{\Lambda}^{L} dr \frac{r^{d-1}}{r^{d+z-2+\eta}}
  \propto L^{2-(z+\eta)}
\end{equation}
with $\Lambda$ being a cutoff, and then $m(L)^2\propto L^{-(z+\eta)}$.
Figure \ref{fig:mag_log}(a), which is a log-log plot of the staggered
magnetization against the system size, clearly supports the critical
behavior in the nonmagnetic region, namely, $0.4<J_2\leq 0.5$.  Note
that the region $0.5<J_2\leq 0.6$ does not contradict the behavior
$S(\boldsymbol{q}_{\rm peak}, L)/N_{\mathrm{s}}\sim \int_{0}^{L} dr\,
r^{d-1}\exp[-r/\xi]/N_{\mathrm{s}} \propto 1/L^{2},$ indicating the
exponential decay of the correlation, because the scaling at $J_2=0.55$
and 0.6 in Fig.~\ref{fig:mag_log} is close to $m^2\propto1/L^2$ within
the uncertainty of the estimate of the exponent arising from the
finite-size effect.

We also calculate the spin-spin correlation function defined as
\begin{equation}
 C(\boldsymbol{r}) = \frac{1}{N_{\mathrm{s}}}\sum_{\boldsymbol{r}'}
  \left\langle \boldsymbol{S}_{\boldsymbol{r}}\cdot
  \boldsymbol{S}_{\boldsymbol{r}+\boldsymbol{r}'}
  \right\rangle.
\end{equation}
By considering the effect of the periodic boundary condition, we assume
that the data fit with the following form:
\begin{equation}
 C(\boldsymbol{r}) \propto
  \frac{1}{|r|^{z+\eta}}
  +\sum_{\boldsymbol{n}\neq (0,0)}
  \left( \frac{1}{|\boldsymbol{r}+L\boldsymbol{n}|^{z+\eta}}
  - \frac{1}{|L\boldsymbol{n}|^{z+\eta}}\right),
  \label{eq:spin_fit}
\end{equation}
where $\boldsymbol{n}=(n_x, n_y)$ is an integer vector.  The power-law
decay of the spin-spin correlation function shown in
Fig.~\ref{fig:mag_log}(b) is consistent with the power-law scaling of
the magnetic order parameter.

Therefore, we conclude that the nonmagnetic phase in $0.4<J_2\leq 0.5$
is critical and gapless, and thus the spin gap in this region should be
scaled by Eq.~(\ref{eq:gap_fit2}).  We note again that the phase in
$0.5<J_2\leq 0.6$ is gapped and that the triplet gap should be fitted
with Eq.~(\ref{eq:gap_fit}).  The obtained triplet gap in the
thermodynamic limit is shown in Fig.~\ref{fig:phaseDiag}.

To investigate the possibility of a VBC order, we next
consider the dimer structure function defined as
\begin{equation}
 S_{\mathrm d}(\boldsymbol{q})=\frac{1}{N_{\mathrm{s}}}\sum_{i,j}
  e^{i\boldsymbol{q}\cdot(\boldsymbol{r}_i-\boldsymbol{r}_j)}\left(
  \langle B_i^x B_j^x\rangle
  -\langle B_i^x\rangle\langle B_j^x\rangle\right),
\end{equation}
where $B_i^x$ is a bond operator along the $x$-axis,
$B_i^x=\boldsymbol{S}_i\cdot\boldsymbol{S}_{i+\hat{x}}$.  In the
nonmagnetic region, $S_{\mathrm d}(\boldsymbol{q})$ has a peak at
$\boldsymbol{q}_x=(\pi,0)$, which indicates an expected columnar or
plaquette VBC order.  Therefore, we consider the dimer order parameter
to characterize the VBC phase, defined as $m_{\mathrm d}^2=S_{\mathrm
d}(\boldsymbol{q}_x)/N_{\mathrm{s}}$.

\begin{figure}[h]
 \centering
 \includegraphics[scale=0.5]{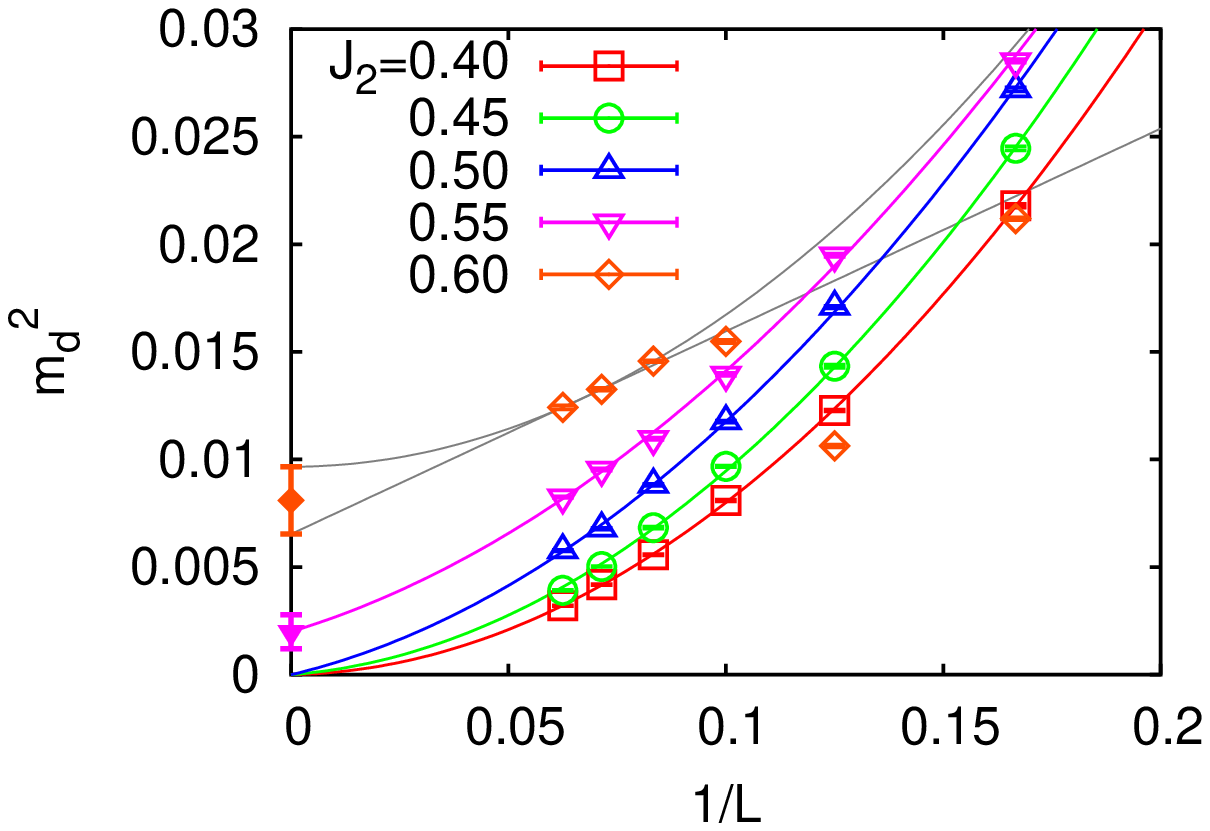}
 \includegraphics[scale=0.5]{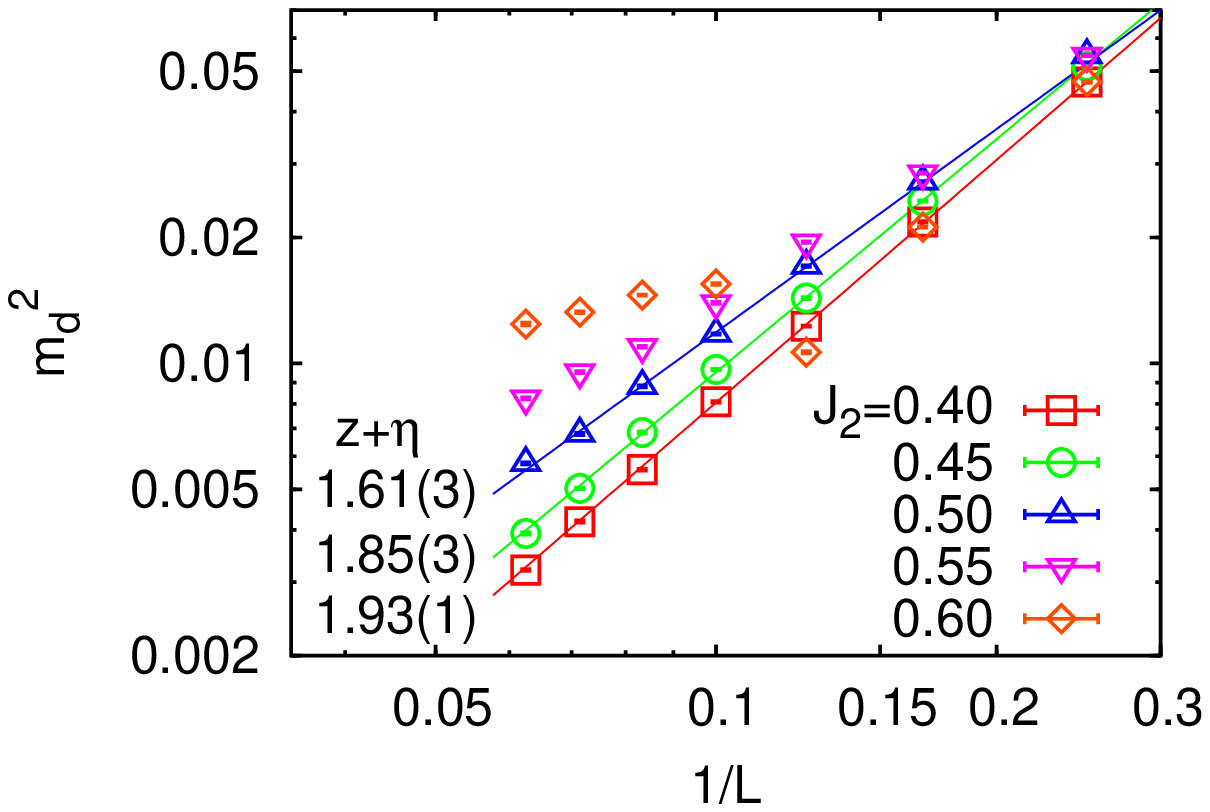}

 \caption{(Color online) Size dependences of dimer order parameters. (a)
 The data are fitted by $a+b/L+c/L^2$. (b) Log-log plot of the same
 data.  The solid lines are obtained by fitting the data with
 $L^{-(z+\eta_{\rm d})}$.}

 \label{fig:dimer_fit}
\end{figure}

\begin{figure}[h]
 \centering
 \includegraphics[scale=0.5]{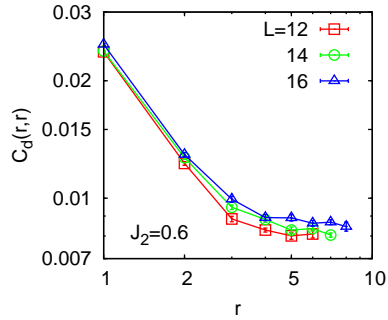}

 \caption{(Color online) Log-log plot of real-space dimer correlation
 function along the diagonal line ($x=y$) at $J_2=0.6$.}

 \label{fig:dr}
\end{figure}

\begin{figure}[h]
 \centering
 \includegraphics[scale=0.5]{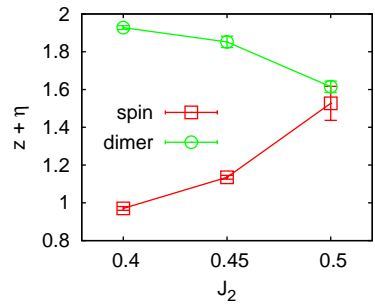}

 \caption{(Color online) Exponent $z+\eta$ estimated from the staggered
 AF magnetic order parameter [Fig.~\ref{fig:mag_log}(a)] and dimer order
 parameter [Fig.~\ref{fig:dimer_fit}(b)].}

 \label{fig:exponent}
\end{figure}

In Fig.~\ref{fig:dimer_fit}, we show the size dependences of the dimer
order parameter $m_{\mathrm d}^2$.  By fitting $m_d^2$ with a quadratic
of $1/L$ in the same way as the magnetic order parameter, we obtain no
VBC order in the gapless region $J_2\leq 0.5$. On the other hand, the
gapped nonmagnetic phase for $0.5<J_2\leq 0.6$ has a small VBC
order. For $J_2=0.6$, since the finite-size effect is strong, we
estimate the upper and lower bounds of the extrapolated value of $m_d^2$
by fitting the last three points from $L=12$ to $16$ with $a+c/L^2$ and
$a'+b'/L$.  We note that these three points are convex as a function of
$1/L$, which implies the finite dimer order parameter in the
thermodynamic limit. The real-space dimer correlation function at
$J_2=0.6$ is shown in Fig.~\ref{fig:dr}, which is consistent with the
long-range ordered state with $m_d^2\sim 0.009$ in agreement with
Fig.~\ref{fig:dimer_fit}(a) .

Figure \ref{fig:dimer_fit}(b) shows a log-log plot of the dimer order
parameter. We find that the data fit well with the power-law scaling
form $m_d^2 \propto L^{-(z+\eta_{\rm d})}$ in the gapless region $0.4<J_2 \leq
0.5$. For $J_2=0.4$, the obtained exponents $z+\eta_{\rm d}$ are close to two,
which is expected in a state without the VBC order where the VBC
correlation decays exponentially at long distances. The obtained exponent
$z+\eta$ for the spin correlation and $z+\eta_{\rm d}$ for the dimer
correlation are plotted in Fig.~\ref{fig:exponent}. It is remarkable
that the exponent $\eta$ appears to vary with $J_2$.

\section{Discussion}
\label{sec:Discussion}

We first comment on the accuracy of our calculations. In a $4\times 4$
system, the trial wave functions of the form given in
Eq.~(\ref{eq:wave_func}) reproduce exact energies for both ground and
excited states.~\cite{prl_dagotto_ed}.  The calculated variance of
energy, $\langle H^2\rangle-\langle H\rangle^2$, is equal to zero, which
indicates that the obtained states are exact eigenstates of the
Hamiltonian.  For a $6\times 6$ lattice, the ground-state energy per
site with $J_2=0.6$ obtained by the mVMC method is
$E/N_{\mathrm{s}}=-0.50355(1)$, while that obtained by exact
diagonalization is $-0.50381$. The error of the energy is one order of
magnitude smaller than the preceding VMC result based on the
projected-BCS state~\cite{arxiv_becca}.  In larger systems, the
ground-state energy of our calculation is comparable to the results
obtained after one Lanczos step reported in Ref.~\citen{prb_hu_vmc}.

We determined the quantum numbers of the ground and excited
states by the projection technique.  In the region $J_2<0.4$, the
results in Figs.~\ref{fig:energy} and \ref{fig:energy_level} are
consistent with the expectation that the lowest-energy triplet state
with $\boldsymbol{K}=(\pi,\pi)$ will become degenerate with the ground state
in the thermodynamic limit as expected in the staggered AF order.
On the other hand, in the region $J_2>0.6$, the results are consistent
with the expectation that the lowest-energy triplet states with
$\boldsymbol{K}=(\pi,0)$ and $(0,\pi)$ together with the
ground state will become degenerate as expected in the stripe AF
order.

In the nonmagnetic region, the lowest excitation energy with each
quantum number is smaller than the staggered and stripe AF phases
(Fig.~\ref{fig:energy_level}). This implies that strong geometric
frustrations destabilize magnetic ordered states.

\begin{figure}[h]
 \centering
 \includegraphics[scale=0.5]{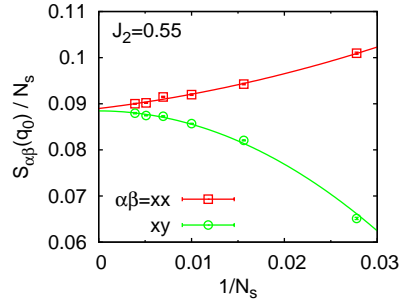}

 \caption{(Color online) Dimer structure factors
 $S_{\alpha\beta}(\boldsymbol{q})$ at $\boldsymbol{q}_0=(0,0)$ for
 $J_2=0.55$. The data are fitted by square polynomials of
 $1/N_{\mathrm{s}}$.}

 \label{fig:bbsq}
\end{figure}

\begin{figure}[h]
 \centering
 \includegraphics[scale=0.5]{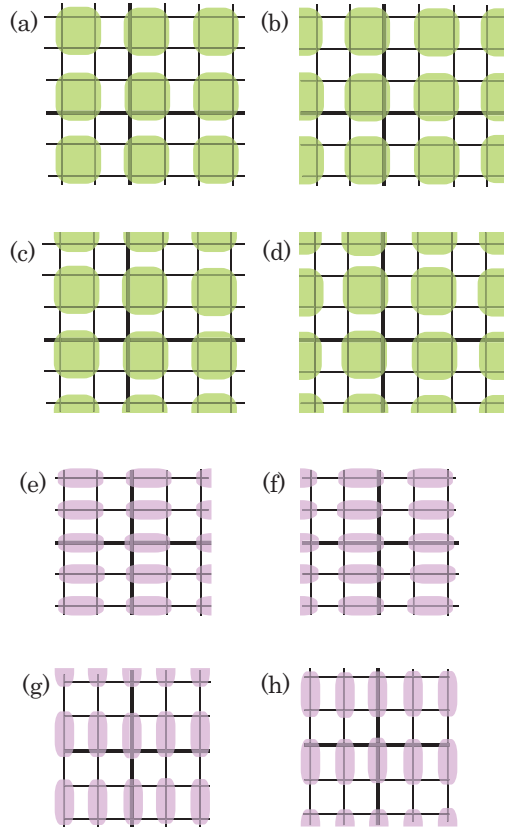}

 \caption{(Color online) (a),(b),(c),(d) Schematic illustration of
 fourfold degenerate plaquette VBC order. The plaquettes are illustrated
 by the shaded green area. (e),(f),(g),(h) Fourfold degenerate columnar
 VBC order. The dimers are illustrated by the shaded purple area. }

 \label{fig:VBC_order}
\end{figure}

The criterion for distinguishing between columnar and plaquette VBC orders was
proposed by Mambrini {\it et al.},~\cite{prb_mambrini_vbc} who
discussed the difference in the dimer structure factors at
$\boldsymbol{q}_0=(0,0)$,
\begin{equation}
 C_\text{col} \simeq \frac{1}{N_{\mathrm{s}}^2}
  \sum_{ij} \langle B_i^x B_j^x - B_i^x B_j^y \rangle
  = \frac{1}{N_{\mathrm{s}}}
  \left[S_{xx}(\boldsymbol{q}_0)- S_{xy}(\boldsymbol{q}_0)\right],
\end{equation}
\begin{equation}
 S_{\alpha\beta}(\boldsymbol{q})
  = \frac{1}{N_{\mathrm{s}}} \sum_{ij}
  e^{i\boldsymbol{q}\cdot(\boldsymbol{r}_i-\boldsymbol{r}_j)}
  \langle B_i^{\alpha} B_j^{\beta} \rangle,
\end{equation}
and argued that $C_\text{col}$ was zero for the plaquette VBC order, but
nonzero for the columnar VBC order, based on the perfectly ordered VBC
state. As shown in Fig.~\ref{fig:bbsq}, the VBC phase satisfies the
criterion $C_\text{col}\rightarrow 0$ in the thermodynamic limit.

However, this criterion is insufficient for distinguishing between the
two VBC phases. A counterexample of their criterion is found in an
imperfect columnar VBC state. Let us assume that, in the state
$|e\rangle$ illustrated in Fig.~\ref{fig:VBC_order}(e), the expectation
value of the bond operator satisfies $B_i^{x}=b_0$ if the bond
$(i,i+\hat{x})$ is on the shaded dimer; otherwise, $B_i^x=b_1$. We also
assume that the bond operators perpendicular to dimers have
$B_i^y=b_2$. The perfectly ordered VBC state satisfies $b_1=b_2=0$. The
other states ($f$, $g$, $h$) are generated by the spatial translations
and the $\pi/2$ rotation. The expectation value of $C_\text{col}$ for
$|e+f+g+h\rangle$ becomes zero if $b_0+b_1=2b_2$, although this state
still has the columnar VBC order. Thus, the criterion
$C_\text{col}\rightarrow 0$ is insufficient to exclude the columnar VBC
order.

The extrapolated values of
$S_{\alpha\beta}(\boldsymbol{q}_0)/N_{\mathrm{s}}$ in
Fig.~\ref{fig:bbsq} and the dimer order parameter $m_d$ in
Fig.~\ref{fig:dimer_fit} impose restrictions on both the columnar and
plaquette VBC orders.  Note that
$m_d^2=S_{xx}(\boldsymbol{q}_x)/N_{\mathrm{s}}$ because of the
translational symmetry.  For the columnar VBC state with $b_0+b_1=2b_2$,
we have $S_{xx}(\boldsymbol{q}_0)/N_{\mathrm{s}} \rightarrow
(b_0+b_1)^2/4$ and $m_d^2 \rightarrow (b_0-b_1)^2/8$.

On the other hand, for the plaquette VBC state illustrated in
Figs.~\ref{fig:VBC_order}(a)-\ref{fig:VBC_order}(d), let us assume that
$B_i^{\alpha}=p_0$ if the bond $(i,i+\hat{\alpha})$ is on the shaded
plaquette; otherwise, $B_i^{\alpha}=p_1$. Then, we obtain
$S_{xx}(\boldsymbol{q}_0)/N_{\mathrm{s}} \rightarrow (p_0+p_1)^2/4$ and
$m_d^2 \rightarrow (p_0-p_1)^2/4$.  For $J_2=0.55$, the finite-size
extrapolations show $S_{xx}(\boldsymbol{q}_0)/N_{\mathrm{s}}=0.0889(4)$
and $m_d^2=0.0020(7)$ in the thermodynamic limit. Therefore, if the
columnar VBC order exists, $b_0=0.36(2)$ and $b_1=0.25(2)$ must be
satisfied, while, if the plaquette VBC order exists, $p_0=0.34(1)$ and
$p_1=0.25(1)$ are required.

\begin{table}[h]
 \centering \caption{Total momenta ($\boldsymbol{K}$) and the
 irreducible representations ($\beta$) of the four singlet states for
 the plaquette and columnar VBC orders illustrated in
 Fig.~\ref{fig:VBC_order}} \label{table:VBC}

 \begin{tabular}{c|cc||c|cc} \hline
  plaquette & $\boldsymbol{K}$ & $\beta$ &
  columnar &  $\boldsymbol{K}$ & $\beta$ \\ \hline
  $a+b+c+d$ & $\Gamma$ $(0,0)$ & $A_1$ & $e+f+g+h$ & $\Gamma$ $(0,0)$ & $A_1$\\
  $a-b+c-d$ & $X$ $(\pi,0)$ & $B_2$ & $e-f$ & $X$ $(\pi,0)$ & $B_2$\\
  $a+b-c-d$ & $X'$ $(0,\pi)$ & $B_1$ & $g-h$ & $X'$ $(0,\pi)$ & $B_1$\\
  $a-b-c+d$ & $M$ $(\pi,\pi)$ & $B_2$ & $e+f-g-h$ & $\Gamma$ $(0,0)$ & $B_1$ \\ \hline
 \end{tabular}
\end{table}

Here, we discuss the plausible VBC order pattern from the viewpoint of
its excitation.  The linear combinations of four states for the
plaquette VBC order illustrated in
Figs.~\ref{fig:VBC_order}(a)-\ref{fig:VBC_order}(d) and the columnar VBC
order in Figs.~\ref{fig:VBC_order}(e)-\ref{fig:VBC_order}(h) can
construct four singlets, whose quantum numbers are listed in
Table~\ref{table:VBC}.  For both of the VBC states, three of the four
singlet states have the same symmetries, namely, $A_1$ at the $\Gamma$
point, $B_2$ at the $X$ point, and $B_1$ at the $X'$ point. As shown in
Fig.~\ref{fig:energy_level}, our VBC state well reproduces this
degeneracy at $J_2=0.55$; the energy of the singlet state with $B_2$ at
the $X$ point (and the equivalent $B_1$ at the $X'$ point) is very close
to the ground-state energy at the $\Gamma$ point with the $A_1$
symmetry.

The only difference between the plaquette and columnar VBC orders
appears in the singlet excitation spectra at the $\Gamma$ and $M$
points.  If the plaquette VBC order is realized, the singlet state with
$B_2$ irreducible representation at the $M$ point should be degenerate
with the other three states in the thermodynamic limit.  However, such a
behavior is not observed in the present calculation
(Fig.~\ref{fig:energy_level}).  On the other hand, the negligible energy
gap between singlet states with $A_1$ and $B_1$ at the $\Gamma$ point is
compatible with the columnar VBC order.

Although our results for excitation spectra support the columnar VBC
phase, results for larger system sizes are desired to decisively
conclude whether the singlet state with $B_2$ at the $M$ point becomes
degenerate with the ground state at the $\Gamma$ point in the
thermodynamic limit.  The plaquette-plaquette correlation allows us to
more directly clarify whether the VBC phase in $0.5<J_2\leq 0.6$ is
columnar or plaquette. However, since an $m$-spin correlation function
requires computational costs scaled by at least $O(m^3)$, calculations
of plaquette-plaquette correlations requiring computations of 8-spin
correlations remain a future challenge.

\begin{figure}[h]
 \centering
 \includegraphics[scale=0.5]{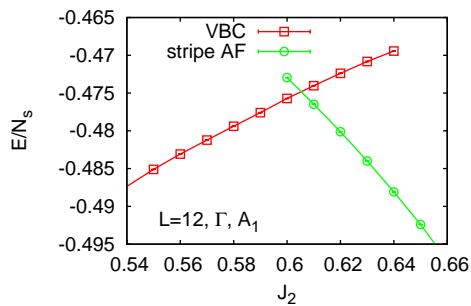}

 \caption{(Color online) Level crossing between the VBC and stripe AF
 states at the first-order transition point at approximately $J_2=0.6$
 for the $12\times 12$ lattice system. Both states have a total momentum
 $\boldsymbol{K}=(0,0)$ and a irreducible representation $A_1$.}

 \label{fig:levelCross}
\end{figure}

Our data support the notion that the first-order transition occurs
between the stripe AF phase and the nonmagnetic region at approximately
$J_2=0.6$. As previously shown, the quantum numbers of the lowest
triplet state change at this transition point.  We also observe that,
for $L=4n$, the metastable state with the same quantum numbers as the
ground state survives around the transition point, as shown
Fig.~\ref{fig:levelCross}.

On the other hand, the phase transition between the staggered AF phase
and the nonmagnetic region at approximately $J_2=0.4$ is continuous
with the critical exponent $\beta\sim1/2$ since the square of the
magnetic order parameter linearly depends on $J_2$ around the phase
boundary [Fig.~\ref{fig:mag_size}(a)].  This behavior was also observed
in previous calculations~\cite{prb_jiang_dmrg,prl_gong_dmrg}.  At
$J_2=0.4$, the staggered AF order is close to the critical point and the
magnetic order parameter fits well with the critical scaling form
$L^{-(z+\eta)}$, as shown in Fig.~\ref{fig:mag_log}. The obtained
exponent $(z+\eta)$ is close to unity. This fact indicates that the
phase transition between the staggered AF and nonmagnetic phases has a
very small $\eta$ with a quantum criticality $z=1$, expected from the
linear dispersion at the critical point by taking $\Delta_{\infty}=0$ in
Eq.~(\ref{eq:dispersion}).  Further studies are necessary to understand
why the obtained critical exponents are close to the mean-field value.
The same characteristic behavior $z+\eta\sim 1$ is also observed in the
$J_1$-$J_2$ Heisenberg model on the triangular
lattice~\cite{arxiv_kaneko_vmc}.

In the gapless region for $0.4<J_2\leq 0.5$, we observed the algebraic
behavior of the spin correlation, which is basically consistent with the
recent DMRG result by Gong {\it et al.},\cite{prl_gong_dmrg} whose
estimated exponent $\eta=0.44$ at $J_2=0.5$ is in agreement with our
result $\eta=0.53(9)$.  As pointed out by Gong {\it et al.}, because the
correlation length can be long, we have some uncertainty in determining
whether this extended gapless region survives in the thermodynamic
limit. If so, the power-law behavior of the spin correlation indicates
that the algebraic spin liquid is realized as a phase in this region. On
the other hand, if this region shrinks to the critical point in the
thermodynamic limit, the deconfined quantum criticality scenario may
become relevant. In this case, our estimated exponents at the transition
points ($\eta\simeq 0.0$ at $J_2=0.4$ and $\eta=0.53(9)$ at $J_2=0.5$)
give the lower and upper bounds of $\eta$, which do not conflict with
$\eta\simeq 0.27$ obtained in the $J$-$Q$ model on the square
lattice~\cite{prb_sandvic_jq}.  However, the size extrapolations of the
staggered as well as the VBC order parameters shown in
Figs.~\ref{fig:mag_fit} and \ref{fig:dimer_fit} strongly suggest that
this is not plausible unless an unknown crossover occurs at larger sizes
beyond the present calculation.

The existence of the gapped phase for $0.5<J_2\leq 0.6$ is contradictory
to the VMC calculation by Hu {\it et al.}~\cite{prb_hu_vmc}.  They
calculated the energy gap between the ground state and the triplet state
at $\boldsymbol{K}=(\pi, 0)$ using the VMC method together with the
Lanczos technique and reported that this gap closes for $J_2>0.48$. We
point out that, since they did not use the quantum number projection
technique, the possibility that their excited states are contaminated by
the ground state cannot be excluded. This causes an underestimation of
the energy gap after the Lanczos steps. In addition, they did not
consider the possibility of the scaling form Eq.~(\ref{eq:gap_fit}).

\section{Conclusions}
\label{sec:Conclusions}

We have calculated the ground and excited states of the spin 1/2
$J_1$-$J_2$ Heisenberg model on the square lattice by the mVMC method
with high accuracy. We emphasize that, in the present study, the
competing phases and their fluctuations can be represented by a unified
framework and the same form of the variational wave function
Eq.~(\ref{eq:wave_func}). The quantum-number projection technique has
allowed us to determine the total momentum and the point group symmetry
for the ground state and excitation structure.  We obtained that the
ground states do not have a magnetic order for $0.4<J_2\leq 0.6$.  By a
careful analysis of the triplet gap and VBC order parameter, we have
found a gapped phase with the VBC order for $0.5<J_2\leq 0.6$.  From the
excitation spectra, we conclude that the VBC order is likely to have the
columnar symmetry, while it does not completely exclude the possibility
of the plaquette order. On the other hand, our data support the
existence of a gapless spin liquid phase for $0.4<J_2\leq 0.5$.  We have
also observed an algebraic behavior of the spin-spin correlation
function, which indicates the realization of the algebraic spin-liquid
state in this region.

In this study, although we focus on a spin system, the mVMC method can
manage itinerant electron systems.  The Hubbard model with the
next-neighbor hopping term connects to the $J_1$-$J_2$ Heisenberg model
in the limit of a large Coulomb repulsion.  It would be intriguing to
investigate in the future how the nature of the spin liquid becomes
modified in the presence of charge fluctuations.

\section*{Acknowledgments}
The mVMC codes used for the present computation are based on that first
developed by Daisuke Tahara. S.M. thanks Takahiro Misawa for fruitful
discussions. This work is financially supported by MEXT HPCI Strategic
Programs for Innovative Research (SPIRE) and Computational Materials
Science Initiative (CMSI). Numerical calculation was partly carried out
at K computer at RIKEN Advanced Institute for Computational Science
(AICS) under grant numbers hp120043, hp120283, hp130007 and hp140215.
Numerical calculation was partly carried out at the Supercomputer
Center, Institute for Solid State Physics, University of Tokyo. This
work was also supported by Grants-in-Aid for Scientific Research
(Nos. 22104010 and 22340090) from MEXT, Japan.

\appendix
\section{Update technique of Pfaffian}

The inner product between the pair wave function and the real-space
electron configuration is given as a Pfaffian of a skew-symmetric
matrix.  The update technique of a Pfaffian for one-electron hopping is
derived on the basis of Cayley's identity~\cite{jpsj_tahara_vmc,
prb_bajdich_pfaffian}.  However, in a spin model, we need to generate
electron configurations by two-electron exchange processes. In this
appendix, we generalize the update technique for $m$-electron move with
an arbitrary number $m$.

First, we summarize the definition and some properties of a Pfaffian.  A
$2n\times 2n$ skew-symmetric matrix $A = [a_{ij}]$ satisfies $A^T =-A$
($a_{ij}=-a_{ji}$), where $A^T$ denotes the transposed matrix of
$A$. The Pfaffian of $A$ is defined as the antisymmetrized product
\begin{equation}
 {\rm Pf} \, A \equiv \mathcal{A}[a_{12}a_{34}\cdots
  a_{2n-1,2n}]
  =\sum_{\alpha} {\rm sgn}(\alpha)
  \prod_{k=1}^n a_{i_k,j_k},
\end{equation}
where the sum runs over all the pair partitions $\alpha=\{(i_1,j_1),
\cdots (i_n,j_n)\}$ with $i_k<i_{k+1}$ and $i_k<j_k$. Here, ${\rm
sgn}(\alpha)$ stands for the parity of the permutation corresponding to
the partition $\alpha$. The Pfaffian satisfies the relations
\begin{equation}
 {\rm Pf}
  \begin{bmatrix}
   A & 0\\
   0 & A'
  \end{bmatrix}
  = {\rm Pf}\,A \times {\rm Pf}\,A',\label{eq:Pf_AA}
\end{equation}
\begin{equation}
 {\rm Pf}[BAB^T] = \det B \times {\rm Pf}\,A,\label{eq:Pf_BAB}
\end{equation}
where $B$ is a $2n \times 2n$ arbitrary matrix.

Our update technique is based on the following identities:
\begin{equation}
 {\rm Pf}\, [A+BCB^T] = {\rm Pf}\, A \times
  \frac{{\rm Pf}\, [C^{-1}+B^T A^{-1} B]}{{\rm Pf}\, [C^{-1}]}
  \label{eq:formula_Pf}
\end{equation}
\begin{equation}
 (A+BCB^T)^{-1} = A^{-1} - A^{-1}B(C^{-1}+B^T A^{-1} B)^{-1} B^T A^{-1},
  \label{eq:formula_inv}
\end{equation}
where we assume that $A$, $B$, and $C$ are a $2n\times 2n$ invertible
skew-symmetric matrix, a $2n\times 2m$ real matrix, and a $2m\times 2m$
invertible skew-symmetric matrix, respectively.  The former identity is
a Pfaffian version of the matrix determinant lemma. The more general
formula of the latter is known as the Woodbury matrix identity.  The
proof of these formulae will be shown later.

To derive the update technique, we focus on a spinless fermion system
for simplicity.  The generalization toward spinful electron systems is
straightforward.  The pair wave function with $2n$ fermions has the form
\begin{equation}
 |\phi\rangle=
  \left(\sum_{r,r'} f_{r,r'}
   c_r^\dagger c_{r'}^\dagger\right)^n
  |0\rangle,
\end{equation}
and the real-space electron configuration is
\begin{equation}
 |x\rangle = c_{r_1}^\dagger c_{r_2}^\dagger \cdots
  c_{r_{2n}}^\dagger |0\rangle,
\end{equation}
where $r_i$ denotes the position of the $i$-th electron.
The commutation relation of fermion operators yields
\begin{equation}
 \langle x | \phi \rangle = n! \, {\rm Pf}\, A,
\end{equation}
where $A$ is a $2n\times 2n$ skew-symmetric matrix with the element
$a_{ij} \equiv f_{r_i,r_j} - f_{r_j,r_i}$.

Suppose that $m$ electrons with indices $\alpha_k$ ($k=1,2,\cdots,m$) in
the electron configuration $|x\rangle$ change their positions from
$r_{\alpha_k}$ to $r'_{\alpha_k}$.  Accordingly, the inner product
between the updated electron configuration $|x'\rangle$ and the pair
wave function is proportional to the Pfaffian of a new skew-symmetric
matrix denoted by $D$.  The matrix $D=[d_{ij}]$ differs from $A$ only in
the $\alpha_k$-th rows and columns.  We assume that the matrices $B$ and
$C$ have the forms
\begin{equation}
 B= \begin{bmatrix}
     U & V
    \end{bmatrix}, \qquad
 C= \begin{bmatrix}
     0 & I\\
     -I & W
    \end{bmatrix},
\end{equation}
where $U$ and $V$ are $2n\times m$ matrices and $W$ is a $m\times m$
skew-symmetric matrix.  If we set the elements of $U$, $V$, and $W$ as
\begin{gather}
 U_{ik} = d_{i,\alpha_k} - a_{i,\alpha_k} \\
 V_{ik} = \delta_{i,\alpha_k} \\
 W_{kl} = -d_{\alpha_k,\alpha_l} + a_{\alpha_k,\alpha_l},
\end{gather}
the desired relation $A+BCB^T=D$ is obtained. The matrix $W$ is
necessary to reduce double counting at $(i,j)=(\alpha_k,\alpha_l)$.
Note that
\begin{equation}
 C^{-1} = \begin{bmatrix}
           W & -I\\
           I & 0
          \end{bmatrix},
\end{equation}
and ${\rm Pf}\,[C^{-1}] = (-1)^{m(m+1)/2}$.

The heaviest part of our update technique is the calculation of $B^T A^{-1}
B$ when $m$ is $O(1)$.  Thus, if we store the inverse matrix of $A$,
the computational cost is $O(n^2)$, while the direct calculation of a Pfaffian
and an inverse matrix requires $O(n^3)$ operations.

Finally, we prove the two identities (\ref{eq:Pf_AA}) and
(\ref{eq:Pf_BAB}), which is easy using the LDU decomposition of a block
matrix, {\it i.e.},
\begin{equation}
 \begin{bmatrix}
  A & B \\
  -B^T & C^{-1}
 \end{bmatrix}
 =
 \begin{bmatrix}
  I & 0 \\
  (A^{-1}B)^T & I
 \end{bmatrix}
 \begin{bmatrix}
  A & 0 \\
  0 & C^{-1}+B^T A^{-1} B
 \end{bmatrix}
 \begin{bmatrix}
  I & A^{-1}B \\
  0 & I
 \end{bmatrix},\label{eq:LDU_decomposition}
\end{equation}
and the UDL decomposition,
\begin{equation}
 \begin{bmatrix}
  A & B \\
  -B^T & C^{-1}
 \end{bmatrix}
 =
 \begin{bmatrix}
  I & BC \\
  0 & I
 \end{bmatrix}
 \begin{bmatrix}
  A+BCB^T & 0 \\
  0 & C^{-1}
 \end{bmatrix}
 \begin{bmatrix}
  I & 0 \\
  (BC)^T & I
 \end{bmatrix}.\label{eq:UDL_decomposition}
\end{equation}
Both decompositions are confirmed by direct calculations of the
right-hand side. We point out that the inverse matrix of $A$ is also
skew-symmetric and that $(A^{-1}B)^T = -B^T A^{-1}$.  Equation
(\ref{eq:formula_Pf}) is derived by taking the Pfaffian of
Eqs.~(\ref{eq:LDU_decomposition}) and (\ref{eq:UDL_decomposition}) and
using properties of the Pfaffian Eqs.~(\ref{eq:Pf_AA}) and
(\ref{eq:Pf_BAB}).  If we take inverse of
Eqs. (\ref{eq:LDU_decomposition}) and (\ref{eq:UDL_decomposition}) and
compare their block elements, we obtain Eq.~(\ref{eq:formula_inv}).

\bibliographystyle{jpsj_mod}
\bibliography{main}

\end{document}